\documentclass {aastex631}

\usepackage{amsmath}

\renewcommand{\vec}[1]{\boldsymbol{#1}}

\shorttitle{Kinetic instabilities in the solar wind}
\shortauthors{Opie et al.}
\begin{document}

\title{Conditions for proton temperature anisotropy to drive instabilities in the solar wind\footnote{Draft}}

\author[0000-0002-2280-8807]{Simon Opie}
\affiliation{Mullard Space Science Laboratory, University College London, Dorking, RH5~6NT, UK }

\author[0000-0002-0497-1096]{Daniel Verscharen}
\affiliation{Mullard Space Science Laboratory, University College London, Dorking, RH5~6NT, UK }

\author[0000-0003-4529-3620]{Christopher H. K. Chen}
\affiliation{Department of Physics and Astronomy, Queen Mary University of London, London E1~4NS, UK}

\author{Christopher J.~Owen}
\affiliation{Mullard Space Science Laboratory, University College London, Dorking, RH5~6NT, UK }

\author{Philip A.~Isenberg}
\affiliation{Space Science Center, University of New Hampshire, Durham NH 03824, USA}



\begin{abstract}

Using high-resolution data from Solar Orbiter, we investigate the plasma conditions necessary for the proton temperature anisotropy driven mirror-mode and oblique firehose instabilities to occur in the solar wind. We find that the unstable plasma exhibits dependencies on the angle between the direction of the magnetic field and the bulk solar wind velocity which cannot be explained by the double-adiabatic expansion of the solar wind alone. The angle dependencies suggest that perpendicular heating in Alfv\'enic wind may be responsible. We quantify the occurrence rate of the two instabilities as a function of the length of unstable intervals as they are convected over the spacecraft. This analysis indicates that mirror-mode and oblique firehose instabilities require a spatial interval of length greater than 2 to 3 unstable wavelengths in order to relax the plasma into a marginally stable state and thus closer to thermodynamic equilibrium in the solar wind. Our analysis suggests that the conditions for these instabilities to act effectively vary locally on scales much shorter than the correlation length of solar wind turbulence.         

\end{abstract}



\section{Introduction} \label{sec:intro}

The solar wind is a continuous stream of plasma from the Sun which exhibits significant measurable variability in its characteristic properties on a range of spatial and temporal scales \citep[for recent reviews, see][]{matteini_ion_2012, bruno_solar_2013,chen_recent_2016,verscharen_multi-scale_2019}. The fundamental processes that heat and accelerate the solar wind are not at present fully understood \citep{tu_mhd_1995, parker_dynamical_1965, cranmer_role_2015}. 

A turbulent cascade is generally invoked to explain how energy injected near the Sun into the solar wind at large scales is transferred to kinetic scales where it is available to heat and accelerate individual particles as the plasma travels radially outwards in a practically collisionless environment \citep{bruno_solar_2013, kiyani_dissipation_2015, alexandrova_solar_2013, chandran_incorporating_2011}. At kinetic scales, a secular energy transfer from electromagnetic field fluctuations into the particles ultimately increases entropy \citep{bale09,chen_multi-species_2016,verscharen16}. In addition, energy transfer from the particles into the electromagnetic fields is possible when free energy in the form of  temperature anisotropy or other non-equilibrium particle features is available. This transfer occurs in the form of  instabilities that lead to characteristic wave--particle interactions. Micro-instabilities act to restore thermodynamic equilibrium in the solar wind, thereby lowering the driving free energy \citep{kunz_firehose_2014, verscharen17, chen_multi-species_2016}. In this way, micro-instabilities play an important role for the macro-scale energy distribution in the solar wind \citep{verscharen_multi-scale_2019}.

The solar wind is often studied in a parameter space defined by the plasma-$\beta$, given by the ratio of plasma pressure to the magnetic pressure, and the ratio between the temperature $T_{\perp}$ perpendicular to the magnetic field and the temperature $T_{\parallel}$ parallel to the magnetic field \citep{gary_proton_2001,kasper_windswe_2002,hellinger_solar_2006,bale09}. We refer to plots of the distribution of the data in this space for any given species in the solar wind as $T_{\perp}/T_{\parallel}$-$\beta$-plot. When contours of parameter combinations reflecting marginal stability to individual unstable modes are added to these  $T_{\perp}/T_{\parallel}$-$\beta$-plots, they demonstrate to what extent the temperature anisotropy is constrained by specific instability modes \citep{chen_multi-species_2016,verscharen16,klein_majority_2018}.

The best-fit constraints to proton temperature anisotropies in $T_{\perp}/T_{\parallel}$-$\beta$ plots are typically provided by the thresholds for the oblique firehose and mirror-mode instabilities \citep{hellinger_solar_2006,bale09, gary_short-wavelength_2015}, which are  non-propagating unstable modes of the Alfv\'en-mode and the slow-mode branches of the dispersion relation \citep{howes_astrophysical_2006,schekochihin_astrophysical_2009,kunz2015,verscharen17}. Sufficient plasma pressure anisotropy creates the necessary conditions for the instabilities to act \citep{chandrasekhar_stability_1958,parker_dynamical_1958, hasegawa_drift_1969, maruca+2012,kunz_firehose_2014,kunz2015}. At large scales they are driven by anisotropies in the total pressure components of all species combined \citep{chen_multi-species_2016}. In this work, we focus solely on the proton contribution to the total pressure anisotropy and the kinetic versions of these instabilities, which create fluctuations on a scale of order the characteristic proton kinetic scales \citep{gary_short-wavelength_2015, howes_dynamical_2015}.

In the case of the oblique firehose instability, excess pressure parallel to the direction of the magnetic field causes the growth of bending in magnetic flux tubes. The magnetic tension force is unable to restore this bending if the pressure anisotropy is sufficiently large. Such a transverse perturbation does not propagate in the form of Alfv\'en waves (as it would in the absence of pressure anisotropy) but grows aperiodically with a polarization similar to Alfv\'en waves \citep{hellinger_oblique_2008,matteini_parallel_2006,kunz_firehose_2014}. For the mirror-mode instability, excess perpendicular pressure leads to the formation of quasi-periodic mirror structures trapping some of the ions between mirror points and setting up compressive standing waves with a wavevector  oblique to the direction of the magnetic field \citep{kivelson1996,Kunz2016,yoon_proton_2021}. Particles accumulate in the region between the mirror points where the magnetic field strength is lower, acting to restore perpendicular pressure balance. This results in the mirror mode being characterised by anti-correlated fluctuations in density and magnetic field strength when seen by a traversing spacecraft \citep[e.g.][]{russell_mirror-mode_1999}. In both cases, the transfer of kinetic particle energy to the electromagnetic fluctuations coincides with a reduction in the anisotropy \citep{Kunz2016, yoon2016}.

Plasma instabilities are usually described in the context of homogeneous and steady-state plasma conditions \citep{gary_theory_1993}. However, the solar wind, like most natural plasmas, is turbulent and thus does not fulfill the assumptions applied in the standard theoretical treatment of these instabilities \citep{kivelson1996,howes_model_2008}. Nevertheless, observations clearly show that instabilities act, at least at some time, in this environment \citep{matteini_ion_2012, maruca+2012, wicks2016, yoon_proton_2021}. Our goal is to quantify the occurrence rates of oblique firehose and mirror-mode  unstable solar wind intervals and their dependence on the direction of the magnetic field. 
We also measure the length of unstable intervals in order to evaluate statistically the spatial homogeneity requirement for these instabilities to effectively reduce the proton temperature anisotropy in the solar wind.
If the occurrence of unstable intervals were determined by a scale-independent process, we would anticipate a smooth and scale-independent statistical distribution of the lengthscales of unstable solar wind intervals. However, if the effective action of the associated instabilities is scale-dependent, we anticipate a break in the statistical distribution of the lengthscales of intervals with unstable plasma parameters. Even without knowing the underlying hypothetical distribution of lengthscales, we conjecture that a break in the statistical distribution into a steeper slope marks the lengthscale above which instabilities are effective. In this interpretation, the homogeneity assumption of linear theory is only sufficiently fulfilled in plasma intervals of length greater than the break scale in the statistical distribution.

\section{Methods} \label{sec:methods}

\subsection{Dataset} \label{subsec:datameth}

Recent space missions have been launched to study the inner heliosphere in great detail, with a focus on the processes that heat and accelerate the solar wind \citep{fox_solar_2016, mueller20, zouganelis2020}. For this study, we use data from Solar Orbiter's Solar Wind Analyser \citep[SWA;][]{owen20} instrument suite, specifically the Proton Alpha Sensor (PAS), and the Magnetometer \citep[MAG;][]{horburymag20}. Solar Orbiter in-situ data are publicly available at the Solar Orbiter Archive\footnote{https://soar.esac.esa.int/soar/}  which is the source for all data in this study. We use data from the cruise phase of the mission in both 2020 and 2021. 

SWA's PAS measures the 3D velocity distribution function (VDF) of protons and $\alpha$-particles, whereby the VDF is assembled over an interval of 1\,s every 4\,s, resulting in a normal-mode cadence of $0.25\,\mathrm{Hz}$ \citep{owen20}. The MAG fluxgate magnetometer provides 8 magnetic-field vectors per second in its normal mode \citep{horburymag20}. We use the PAS proton ground moments data and the MAG normal-mode data in radial, tangential and normal (RTN) coordinates. We average the corresponding MAG vector data over each 1\,s VDF measurement interval from PAS. 

For our statistical analysis, it is convenient to use continuous data intervals of reasonable length. In compiling the full dataset, we select intervals of greater than three consecutive days, subject to data availability.  We only include PAS data with a quality factor $< 0.2$\footnote{According to its definition, data with higher quality are identified with lower quality factor values.} and solar wind bulk velocity $>325\,\mathrm{km}\,\mathrm{s}^{-1}$ with initial selection by visual inspection of the data aided by the SWA-PAS data log\footnote{http://solarorbiter.irap.omp.eu/documents/FEDOROV/}. The intervals chosen are listed in Table~\ref{tab:dataset}. The analysed dataset comprises 975\,516 points in total. No attempt was made to eliminate structures such as shocks, interplanetary coronal mass ejections (ICMEs), or current sheets from the data set.
\begin{deluxetable*}{lCR}
\tablenum{1}
\tablecaption{Data selection from the Solar Orbiter Archive with approximate heliocentric distance for each dataset.\label{tab:dataset}}
\tablewidth{16pt}
\tablehead{
\colhead{Interval} & \colhead{Heliocentric Distance ($R_S$)} & \colhead{Number of Datapoints}
}
%\decimalcolnumbers
\startdata
2020 October 07-18 & 205 & 185\,923 \\
2021 April 22-28 & 190 & 131\,481\\
2021 May 05-11 & 180 & 131\,849\\
2021 June 10-13 & 200 & 79\,641\\
2021 July 06-11 & 190 & 117\,427\\
2021 July 20-24 & 180 & 88\,429\\
2021 October 09-12 & 150 & 81\,362\\
2021 October 19-26 & 160 & 159\,404\\
\enddata

\end{deluxetable*}

We rotate the proton pressure tensor to align with the magnetic field and create a timeseries for $\beta_{\parallel}\equiv 8\pi n_{\mathrm p}k_{\mathrm B}T_{\parallel}/B^2$, where $n_{\mathrm p}$ is the proton number density, $k_{\mathrm B}$ is the Boltzmann constant, and $\vec B$ is the magnetic field averaged over the associated 1\,s PAS measurement interval.  We then also calculate the ratio $T_{\perp}/T_{\parallel}$ for each PAS measurement.

\subsection{Instability thresholds}

We base our analysis on the analytical approximation for the instability thresholds of the anisotropy-driven instabilities in the form
    \begin{equation} \label{ethresh}
        \frac{T_{\perp}}{T_{\parallel}} = 1 + \frac{a}{(\beta_{\parallel} - c)^b},
    \end{equation}
    where $a$, $b$, and $c$ are constants with values given for each instability by \citet{verscharen16}. We use a maximum growth rate of  $\gamma_m  = 10^{-2} \Omega_p$, where $\Omega_p$ is the proton gyrofrequency. We evaluate these instability thresholds for the oblique firehose (OF) and for the mirror-mode (M) instabilities. For reference, we also include the instability thresholds for the  Alfv\'en/ion-cyclotron (A/IC) and fast-magnetosonic/whistler (FM/W) instabilities in part of our analysis.

\subsection{Angle analysis} \label{subsec:angmeth}

Working in RTN coordinates, we calculate the angle between $\vec B$ and $\vec V$ using the complete 3D vectors as 
\begin{equation} \label{eang}
    \theta_{BV}^{\prime} = \text{arccos}\frac{\vec B\cdot \vec V}{BV}, 
\end{equation}
where $\vec V$ is the bulk velocity of the protons. 
We convert the angle $\theta_{BV}^{\prime}$ into a full 360$^\circ$ distribution in order to capture the full range of variability in the fluctuations of the magnetic field and to retain the separation of the sector structure of the solar wind. For this conversion, we define $b=B_R+iB_T$ and $v=V_R+iV_T$.
We then calculate the angle  $\phi_v=\mathrm{arg}(v)$, where $\mathrm{arg}(\cdot) \in [0,2\pi)$ is the polar angle in the complex plane. After rotating $b$ by $-\phi_v$ in the complex plane, we define the difference angle between $b$ and $v$ as $\phi_{bv}=180^{\circ}\mathrm{arg}(be^{-i\phi_v})/\pi$. If $0< \phi_{bv}\le 180^{\circ}$, we set $\theta_{BV}=360^{\circ}-\theta_{BV}^{\prime}$. Otherwise,  we set $\theta_{BV}=\theta_{BV}^{\prime}$.
This procedure leads to a representation of the angle $\theta_{BV}$ between $\vec B$ and $\vec V$ within the range $[0^\circ, 360^\circ)$.

The angle $\theta_{BV}$ is the most appropriate measure for quantifying the fluctuations of $\vec B$ and $\vec V$ within structures convected over the spacecraft as a single point of measurement \citep[see also][and references therein]{woodham21}. This link to the convection speed $\vec V$ is particularly important when Taylor's hypothesis is used to map temporal to spatial data \citep{taylor_spectrum_1938,treumann_applicability_2019}. On average and for large datasets, we expect that $\theta_{BV} \approx \theta_{BR}$, where $\theta_{BR}$ is the angle between $\vec B$ and the unit vector $\hat {\vec R}$ in the radial direction. In the case of $\vec V\propto \hat {\vec R}$,  $\theta_{BV}$ represents the azimuthal angle of $\vec B$ and statistically approaches the Parker angle \citep{parker_dynamical_1965}.   

\subsection{Lengthscale analysis} \label{subsec:lenmeth}

%(explanatory sketch as figure 1)
\begin{figure}[ht!]
\plotone{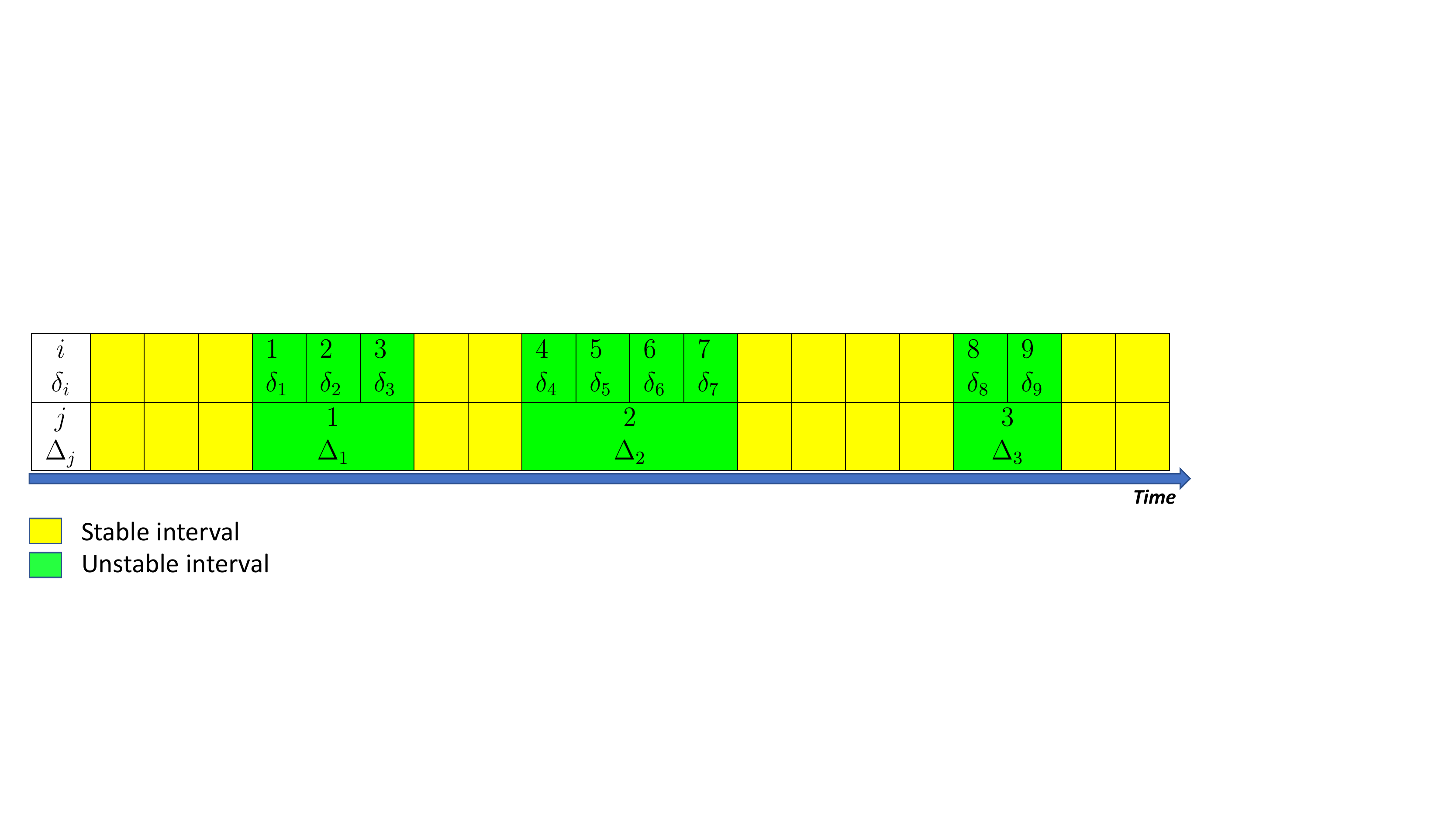}{}
\caption{We identify unstable intervals and calculate their persistence interval ($\Delta_{j}$) by aggregating the lengthscales ($\delta_{i}$) of each measurement interval $i$ in appropriate dimensionless units, where $\delta_{i} = V_i\tau/\rho_{\mathrm pi}$, $ V_i$ is the bulk velocity in interval $i$, $\rho_{\mathrm pi}$ is the  proton gyroradius in interval $i$, and $\tau$ is the interval duration ($4\,\mathrm{s}$). We also calculate $\delta_i$ in units of inertial length ($\mathrm{d_{pi}}$) where $\rho_{\mathrm pi}$ is replaced by $\mathrm{d_{pi}}$. We label sets of consecutive unstable intervals with the number $j$ and calculate their length $\Delta_j$ with Eqs.~(\ref{Delta_rho}) and (\ref{Delta_d}). \label{fig:sketch}}
\end{figure}

We calculate the lengthscales associated with the persistence of instabilities in the solar wind using Taylor's hypothesis \citep{taylor_spectrum_1938}. 
As indicated in Figure~\ref{fig:sketch}, we identify all intervals with parameters above an instability threshold from Eq.~(\ref{ethresh}) in the complete dataset separately for oblique firehose and mirror-mode instabilities. Unstable intervals are shown as individual green boxes in Figure~\ref{fig:sketch}. We calculate the lengthscale $l_i=V_i\tau$ for each unstable interval $i$, where $V_i$ is the proton bulk velocity of interval $i$ and $\tau=4\,\mathrm s$ is the cadence of PAS. Using the proton gyroradius $\rho_{\mathrm pi}$ and the inertial length $d_{\mathrm pi}$ for each individual interval $i$, we then calculate the dimensionless lengthscales $\delta_i^\rho=l_i/\rho_{\mathrm pi}$ and $\delta_i^d=l_i/d_{\mathrm pi}$. The sums of the consecutive dimensionless lengthscales give the total dimensionless persistance interval for each occurrence of the respective instability as convected over the spacecraft. We define them as
\begin{equation}\label{Delta_rho}
    \Delta_j^\rho=\sum\limits_i \delta_i^\rho
\end{equation}
and
\begin{equation}\label{Delta_d}
    \Delta_j^d=\sum\limits_i \delta_i^d,
\end{equation}
where the index $j$ indicates each set of consecutive unstable intervals, and the index $i$ sums over all individual intervals that contribute to the persistence interval $j$.

\section{Results} \label{sec:results}

\subsection{Data overview}

\begin{figure}
\plotone{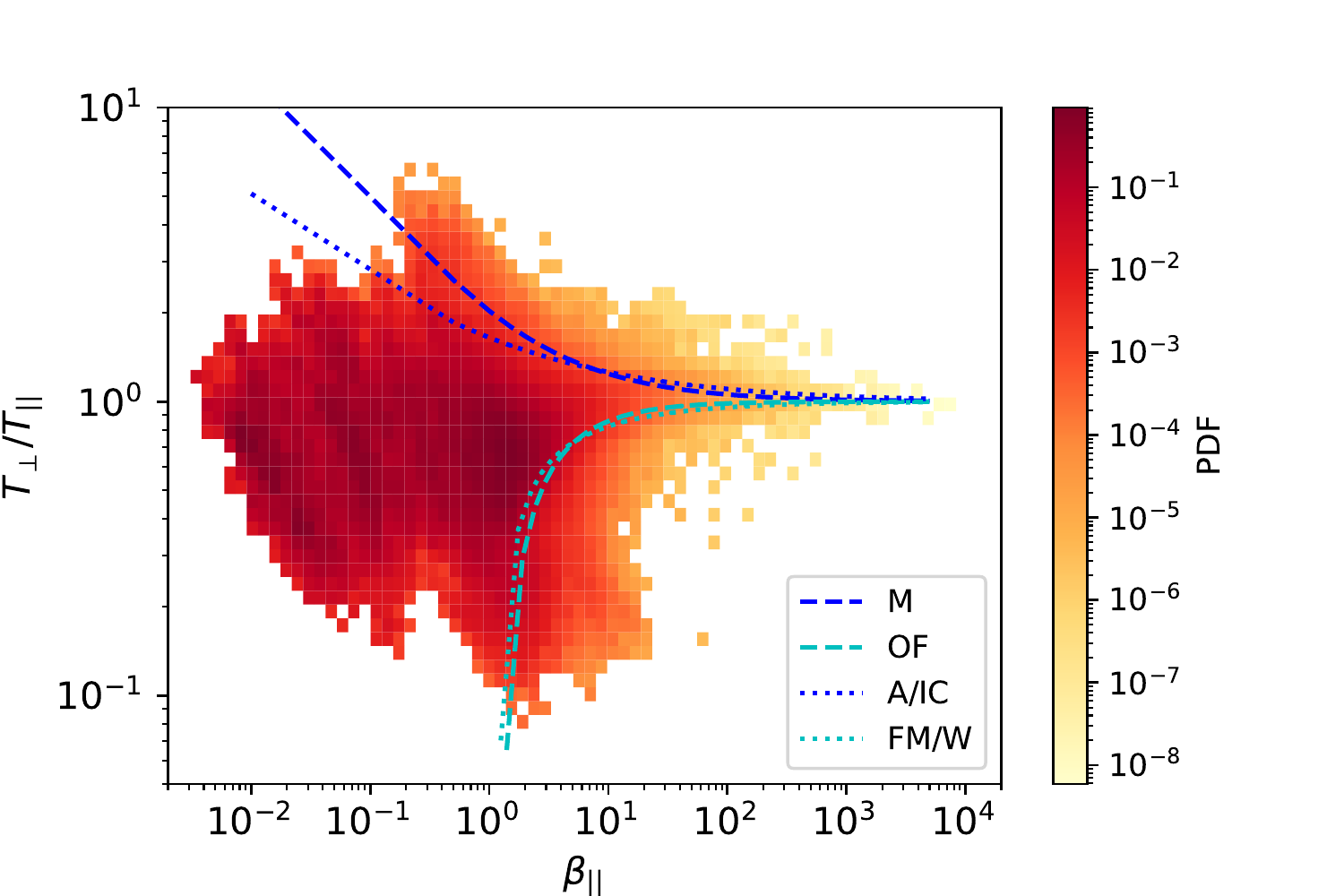}
\caption{Data distribution in a $T_\perp/T_{\parallel}$-$\beta_{\parallel}$-plot. We overplot the instability thresholds according to Eq.~(\ref{ethresh}) for $\gamma_m  = 10^{-2} \Omega_p$. The color-coding indicates the  probability density of datapoints in each bin.  \label{fig:bp}}
\end{figure}

In Figure~\ref{fig:bp}, we show the $T_{\perp}/T_{\parallel}$-$\beta_{\parallel}$-plot of the  probability density function (PDF) of our full dataset. From the total dataset, 940\,598 individual data points are stable to both the mirror-mode and oblique firehose instabilities, within the regime that we classify as both mirror-stable and firehose-stable. 4526 individual data points are in the mirror-mode unstable region ($0.46\%$) and 30\,392 individual data points in the oblique firehose unstable region ($3.12\%$). The instability thresholds  apparently bound the probability distribution, as has been noted by others \citep{hellinger_solar_2006,bale09}. 

The number  of data points in the mirror-mode and oblique firehose unstable regimes is sufficient to allow a separate statistical analysis of these regions. For this investigation, we define four categories of data: 'all' data represents the complete dataset; 'stable' refers to the data points that are stable to both the mirror-mode  and the oblique firehose instabilities; 'oblique firehose unstable' and 'mirror-mode unstable' refers to the data points in the regions beyond their respective instability thresholds with $\gamma_m  > 10^{-2} \Omega_p$.

\subsection{Angle analysis}

\begin{figure*}
\gridline{\fig{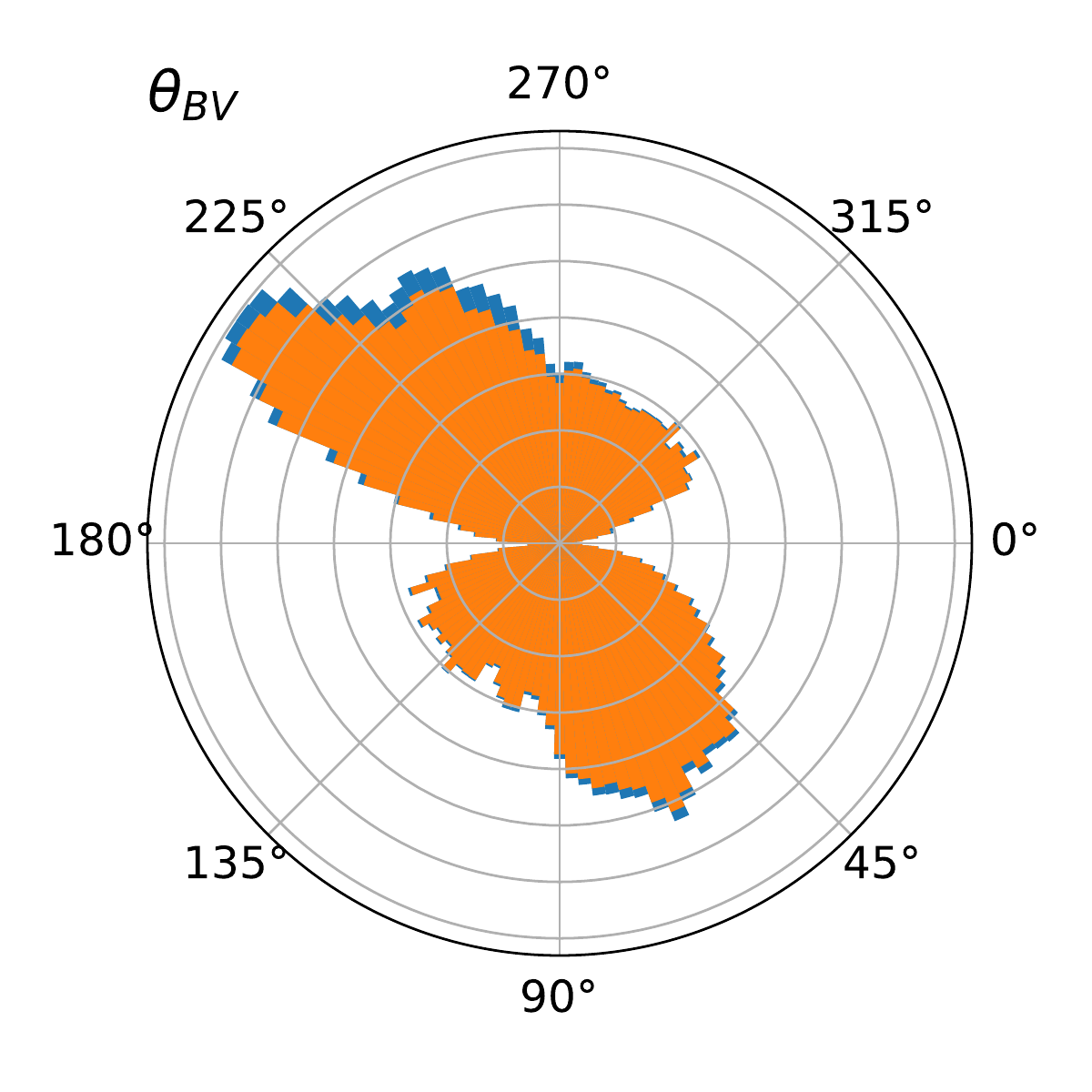}{0.35\textwidth}{(a)}
          \fig{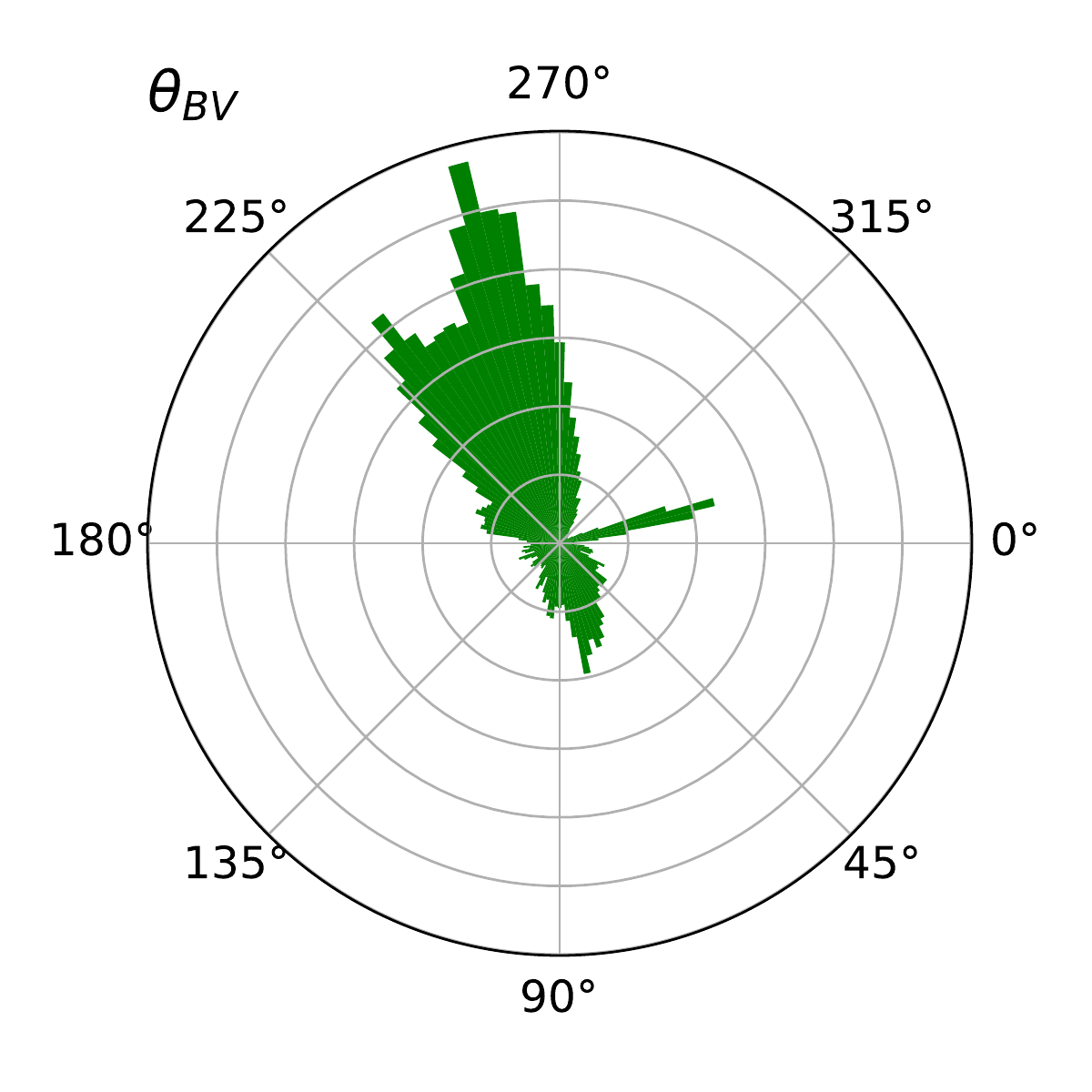}{0.35\textwidth}{(b)}
          \fig{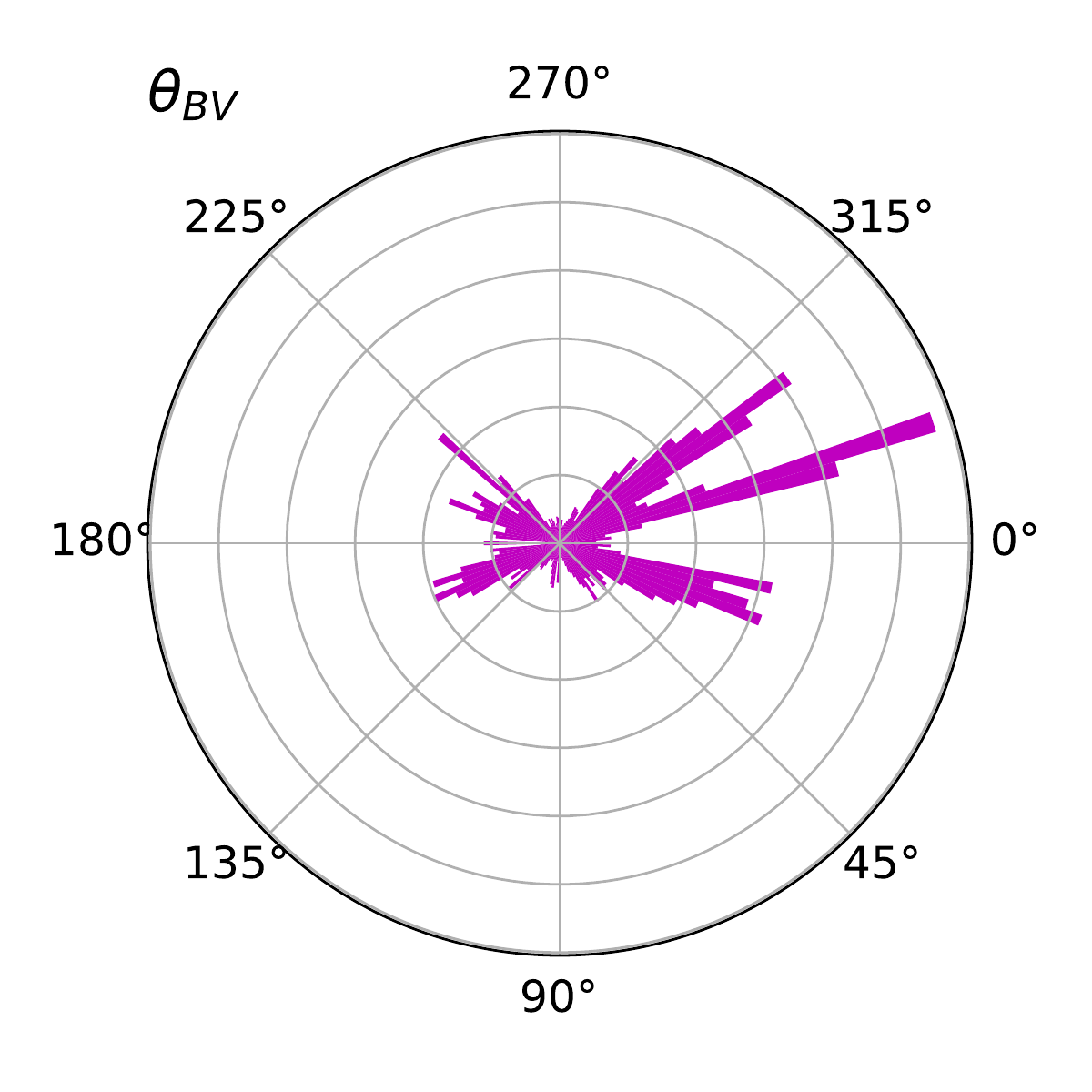}{0.35\textwidth}{(c)}
          }
\caption{Polar plots of the distributions of the four categories of data points as functions of $\theta_{BV}$. The three panels show (a) all data (blue) with stable data (orange) overlaid, (b) oblique firehose unstable data (green), and (c) mirror-mode unstable data (magenta). Both (b) and (c) are normalized according to Eq.~(\ref{eno2}). The grey concentric circles are spaced in increments of (a) 2500, (b) 0.02, and (c) 0.005, starting from zero. $0^\circ$ is the direction of $\vec V$. The distributions are binned at $3^\circ$ resolution.
\label{fig:angles}}
\end{figure*}

\begin{figure}
    \centering
    \includegraphics{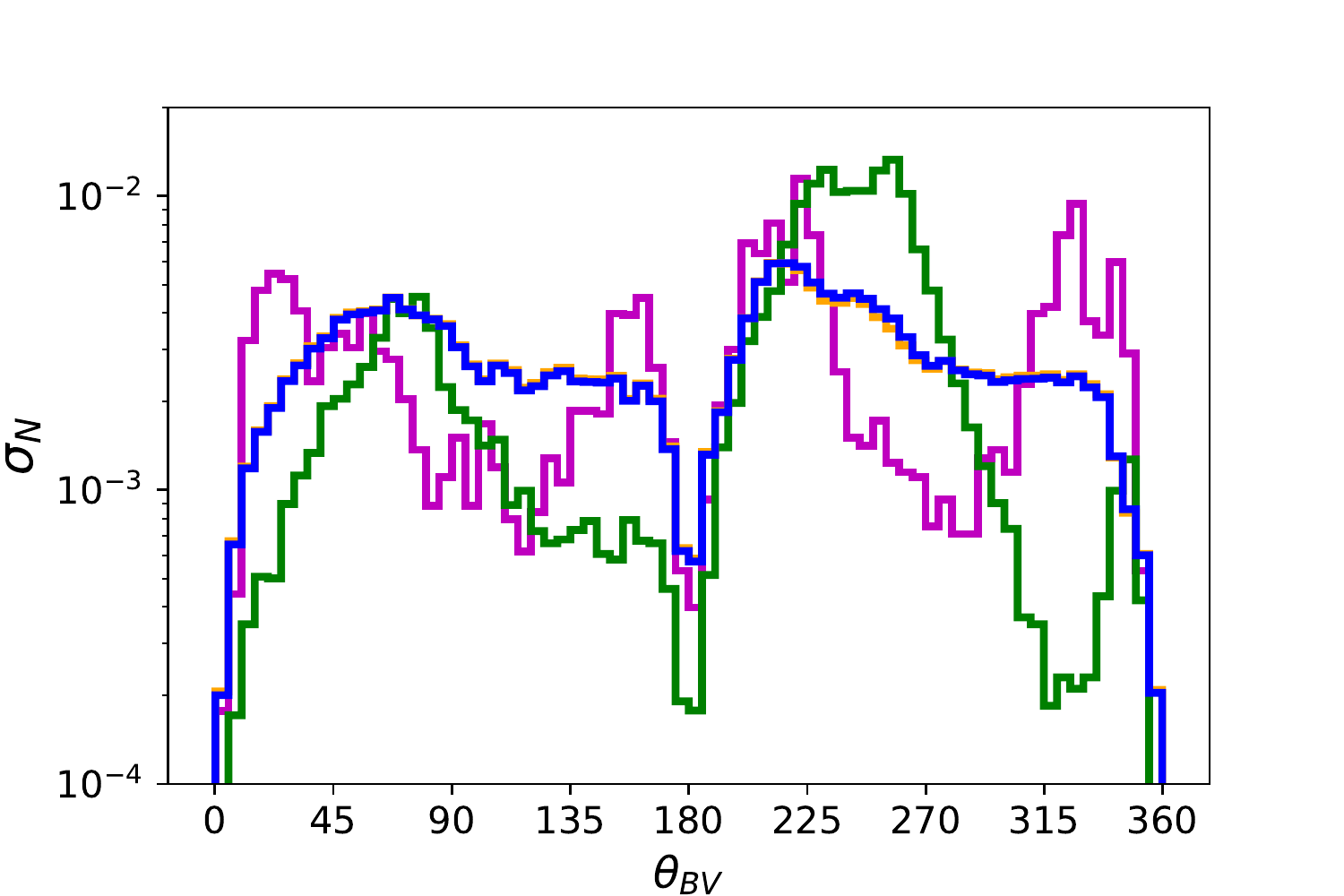}
    \caption{PDF of the angular distribution of our four data categories: all data (blue), stable (orange), oblique firehose unstable (green), mirror-mode unstable (magenta). The orange histogram is mostly obscured by the blue. PDFs, calculated according to Eq.~(\ref{eno}), are binned at $5^\circ$ resolution.
    \label{fig:angles2}}
\end{figure}

Figure~\ref{fig:angles} shows by category of data points (all, stable, oblique firehose unstable, mirror-mode unstable), the distributions of $\theta_{BV}$, calculated as described in Section~\ref{subsec:angmeth}. We quantify the rate of occurrence of unstable data for oblique firehose and mirror-mode by normalizing the distributions by the total number of occurrences of the whole dataset in each angle bin. This normalization quantifies the statistical significance of the excess of unstable modes in each angle bin. We calculate the normalized distribution bin count 
    \begin{equation} \label{eno2}
        \sigma_{ND} = \frac{\sigma_R}{\sigma_{TD}},
    \end{equation}
where  $\sigma_R$ is the bin count of unstable intervals and $\sigma_{TD}$ is the total bin count of the whole dataset for a given angle. The resulting polar plots represent the conformal projection of the 3D angle distribution onto a 2D (RT) plane.
    
Panels (b) and (c) of Figure~\ref{fig:angles} show a clear differentiation in the distribution of the data points that are oblique firehose and mirror-mode unstable. We find that oblique firehose unstable data points occur predominantly when $225^{\circ}\lesssim\theta_{BV}\lesssim 270^{\circ}$. The mirror-mode unstable data points occur predominantly when $\vec B$ and $\vec V$ are within $\sim 45^{\circ}$ of alignment or anti-alignment. 

We note the uneven distribution of the direction of $\vec B$ between the sunward and anti-sunward sectors in our dataset. Data in the upper left quadrants of the plots, where $180^\circ \leq \theta_{BV} \leq 270^\circ$, predominates. We attribute this asymmetry to the position of the spacecraft relative to the current sheet under the quiet solar wind conditions during our data collection period. 

Figure~\ref{fig:angles2} shows by category of data points (all, stable, oblique firehose unstable, mirror-mode unstable), the probability densities of $\theta_{BV}$, calculated as described in Section~\ref{subsec:angmeth}. For purposes of comparison, we plot the normalized density bin count
    \begin{equation} \label{eno}
        \sigma_N = \frac{\sigma_R}{\sigma_T W_b}
    \end{equation}
for each category, where  $\sigma_R$ is the bin count, $\sigma_T$ is the total bin count of the plotted dataset across all angle bins, and $W_b$ is the bin width.
This distribution is normalized  so that $\sum(\sigma_N W_b) = 1$. The maxima of the PDF, shown by the peak values for the distribution of all data are at $\theta_{BV} \approx 70^\circ$ in the anti-sunward and $\approx 225^\circ$ in the sunward direction. The PDF of mirror-mode unstable points peaks and exceeds the PDF of stable points at $\theta_{BV} \approx 25^\circ$, $155^\circ$, $225^\circ$, and $330^\circ$, whereas the PDF of oblique firehose unstable points peaks and exceeds the PDF of stable points at $\theta_{BV} \approx 75^\circ$ and $255^\circ$.

\begin{figure*}
\gridline{\fig{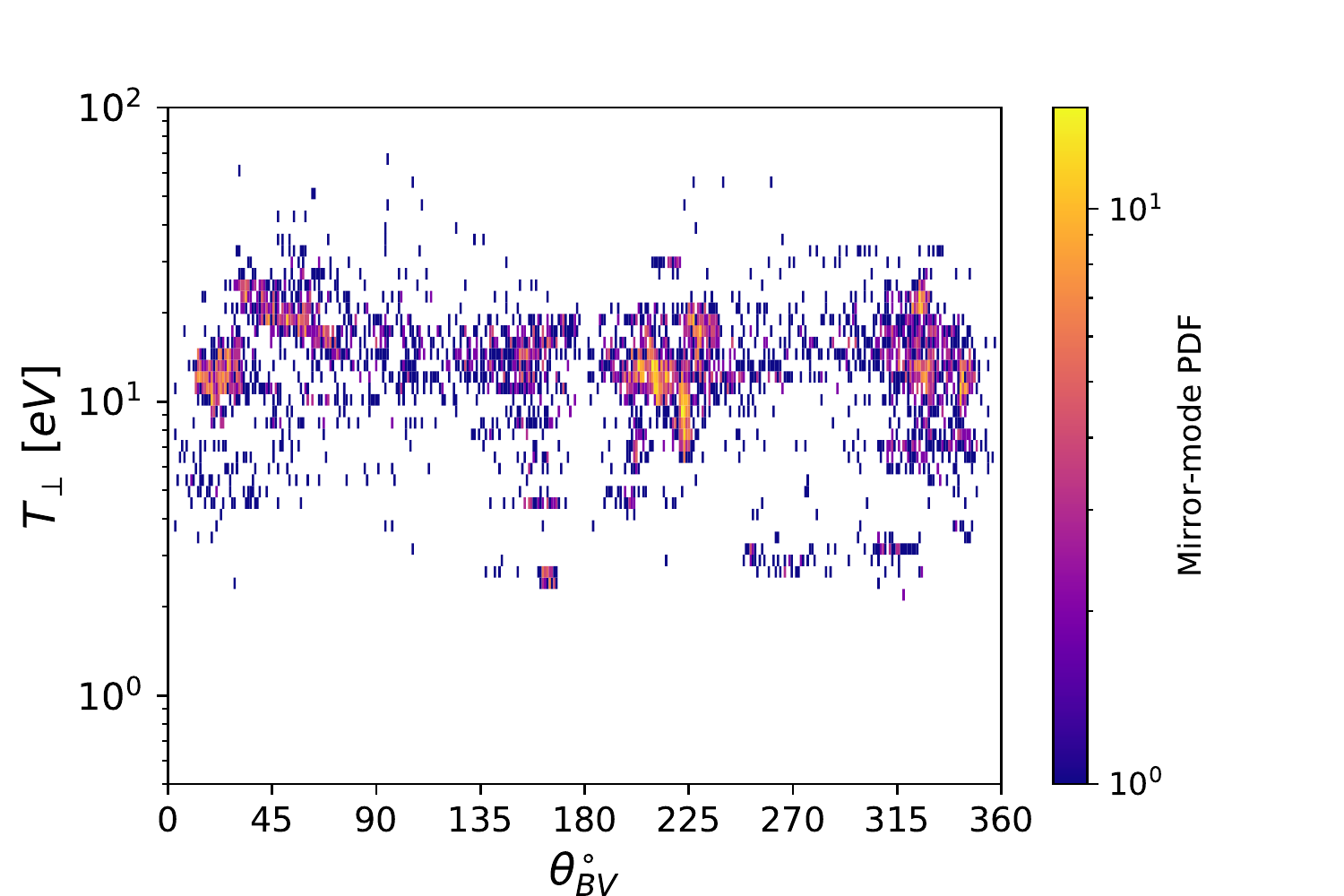}{0.5\textwidth}{(a)}
          \fig{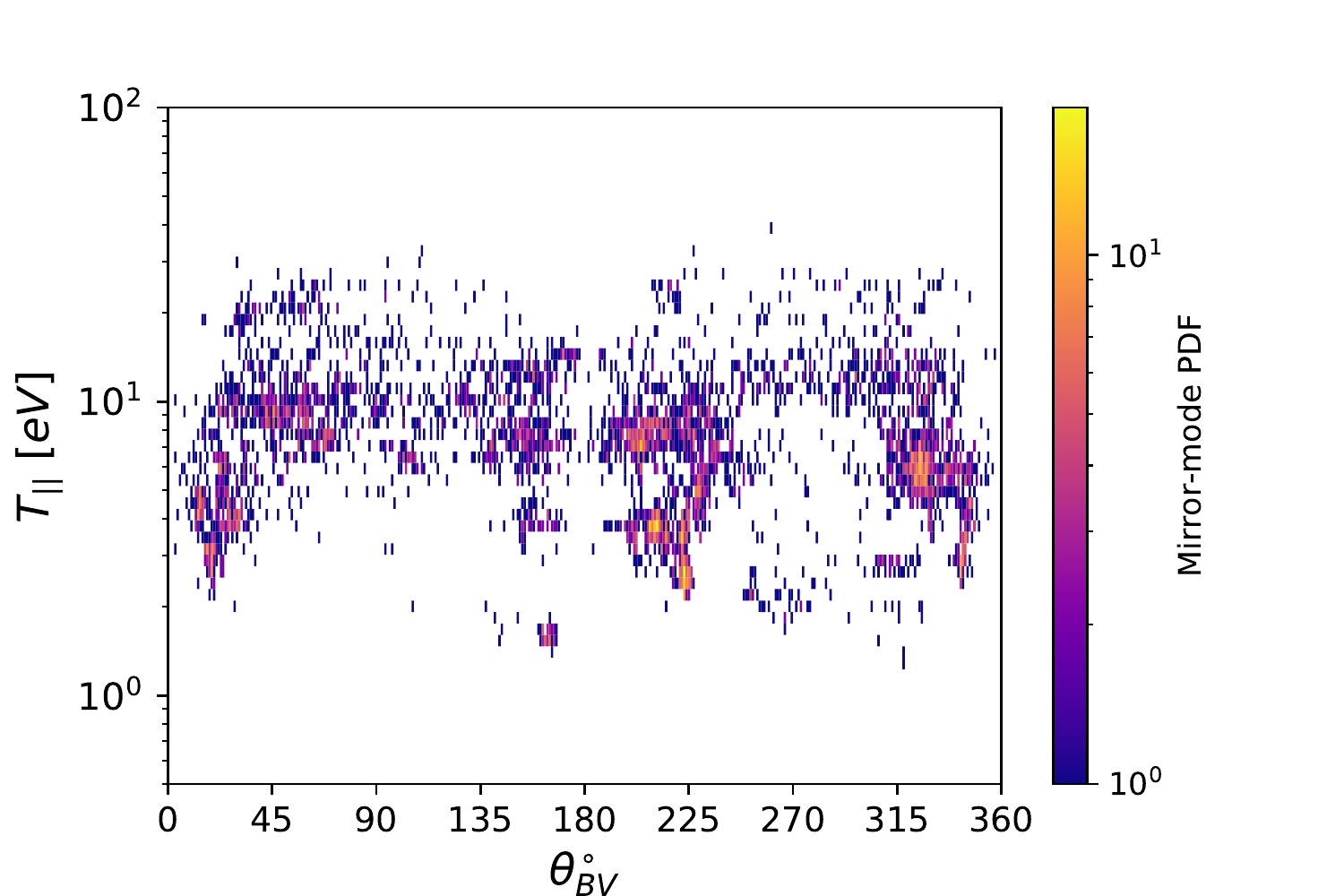}{0.5\textwidth}{(b)}
          }
\gridline{\fig{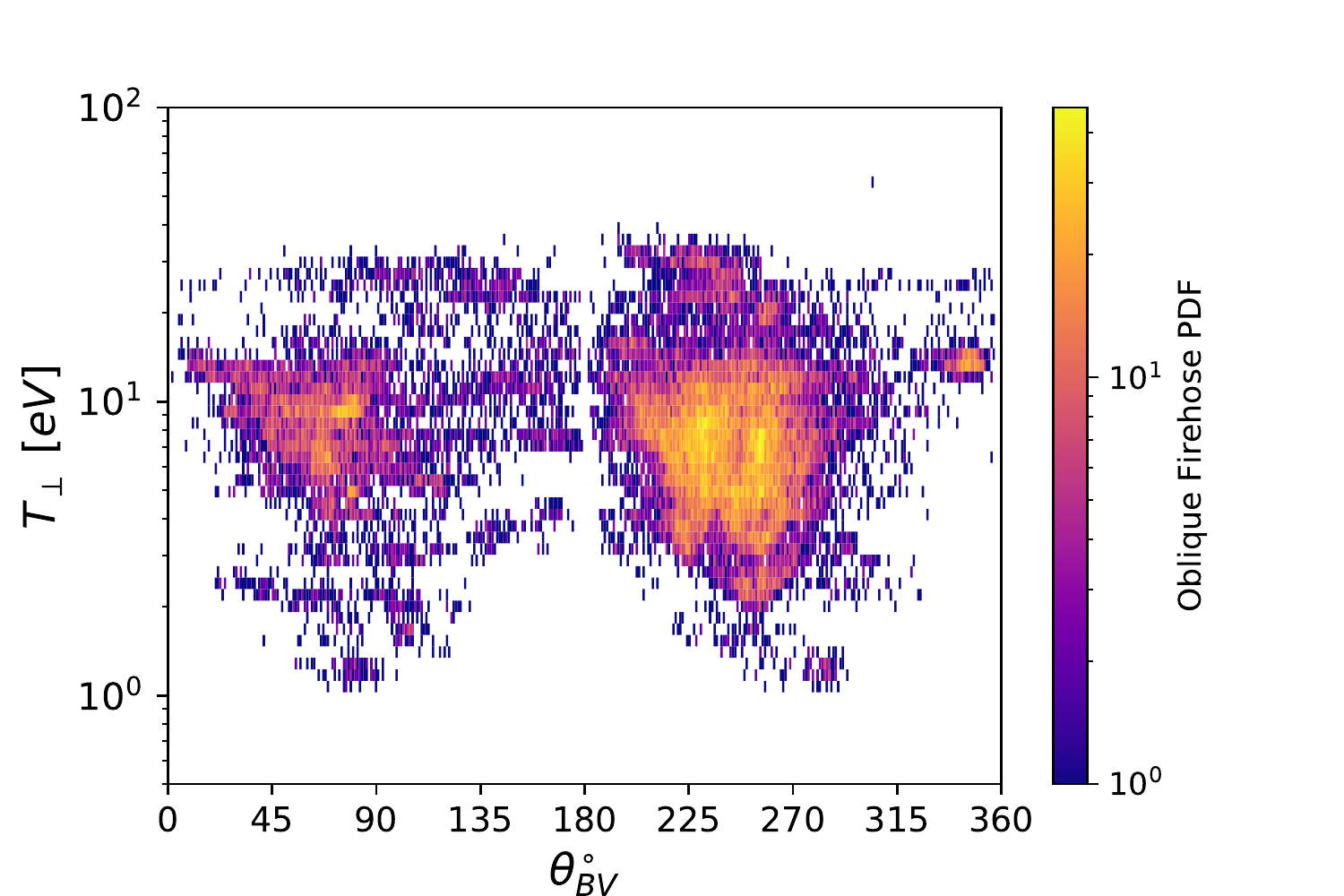}{0.5\textwidth}{(c)}
          \fig{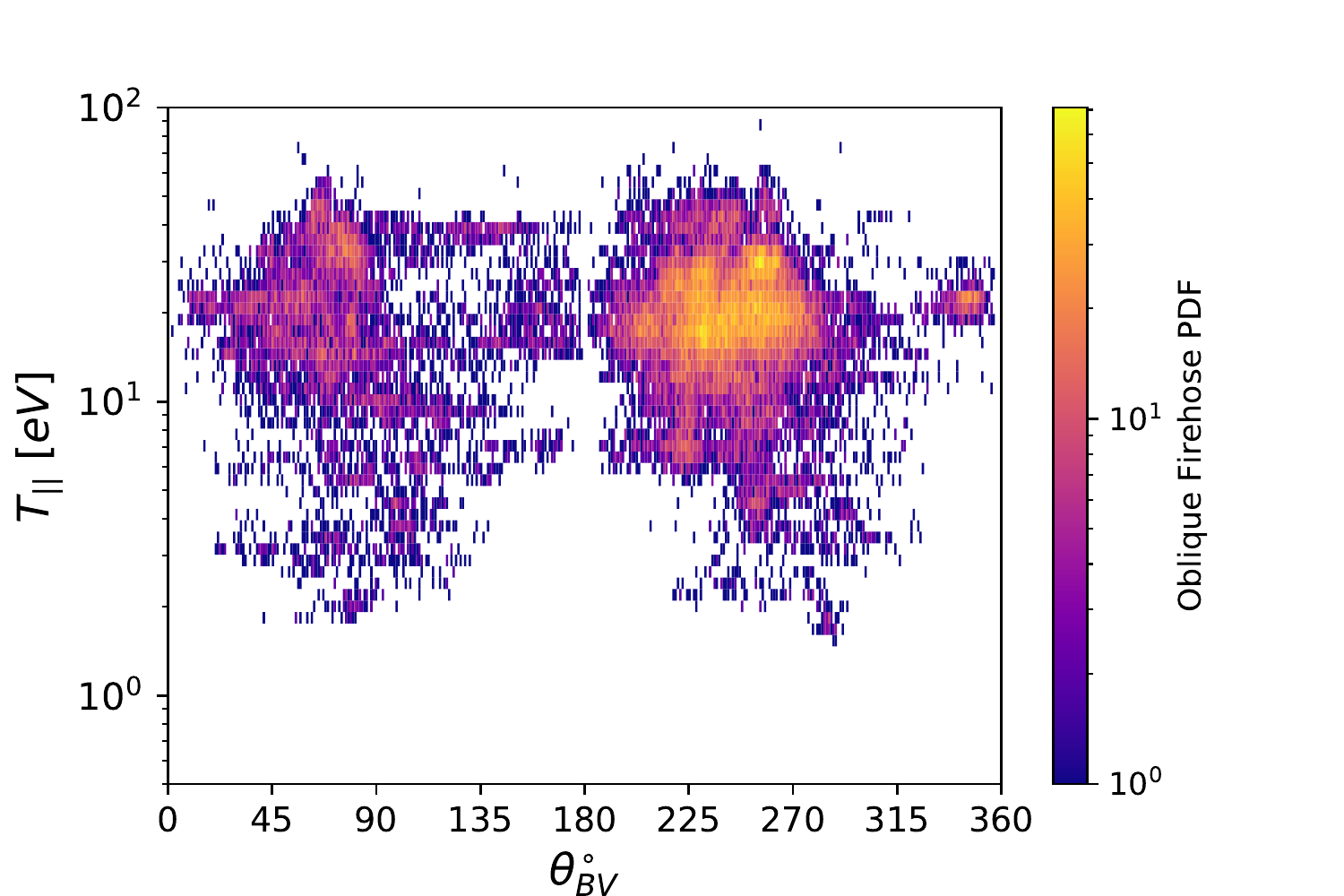}{0.5\textwidth}{(d)}
          }
\caption{Dependence of $T_\perp$ and $T_{\parallel}$ on $\theta_{BV}$ in the solar wind. The upper panels ((a) $T_\perp$ and (b) $T_{\parallel}$) only include mirror-mode unstable points. The lower  panels ((c) $T_\perp$ and (d) $T_{\parallel}$) only include oblique firehose unstable points. The color scale indicates the number of data points per bin.
\label{fig:temp}}
\end{figure*}

We plot the distributions of $T_\perp$ and $T_{\parallel}$ against $\theta_{BV}$ for datapoints in the mirror-mode and oblique firehose unstable categories in Figure~\ref{fig:temp}. In both cases, the distributions of $T_\perp$ and $T_{\parallel}$ separately exhibit variability with  $\theta_{BV}$ consistent with the pattern of angular dependence in Figure~\ref{fig:angles2}. Similar distributions for $\beta_{\parallel}$ (not shown here) do not reveal a marked dependence on $\theta_{BV}$.

\begin{figure*}
\gridline{\fig{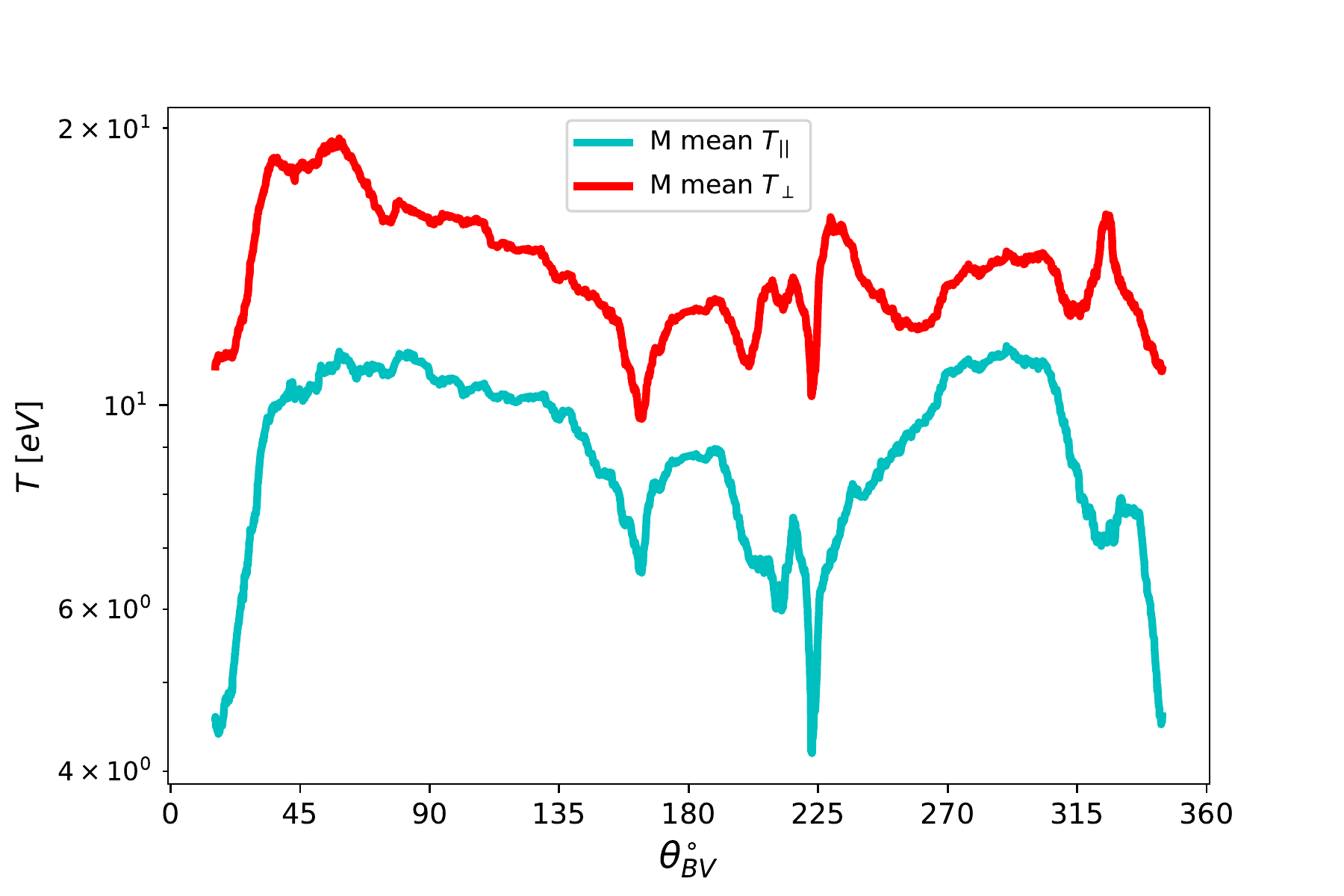}{0.5\textwidth}{(a)}
          \fig{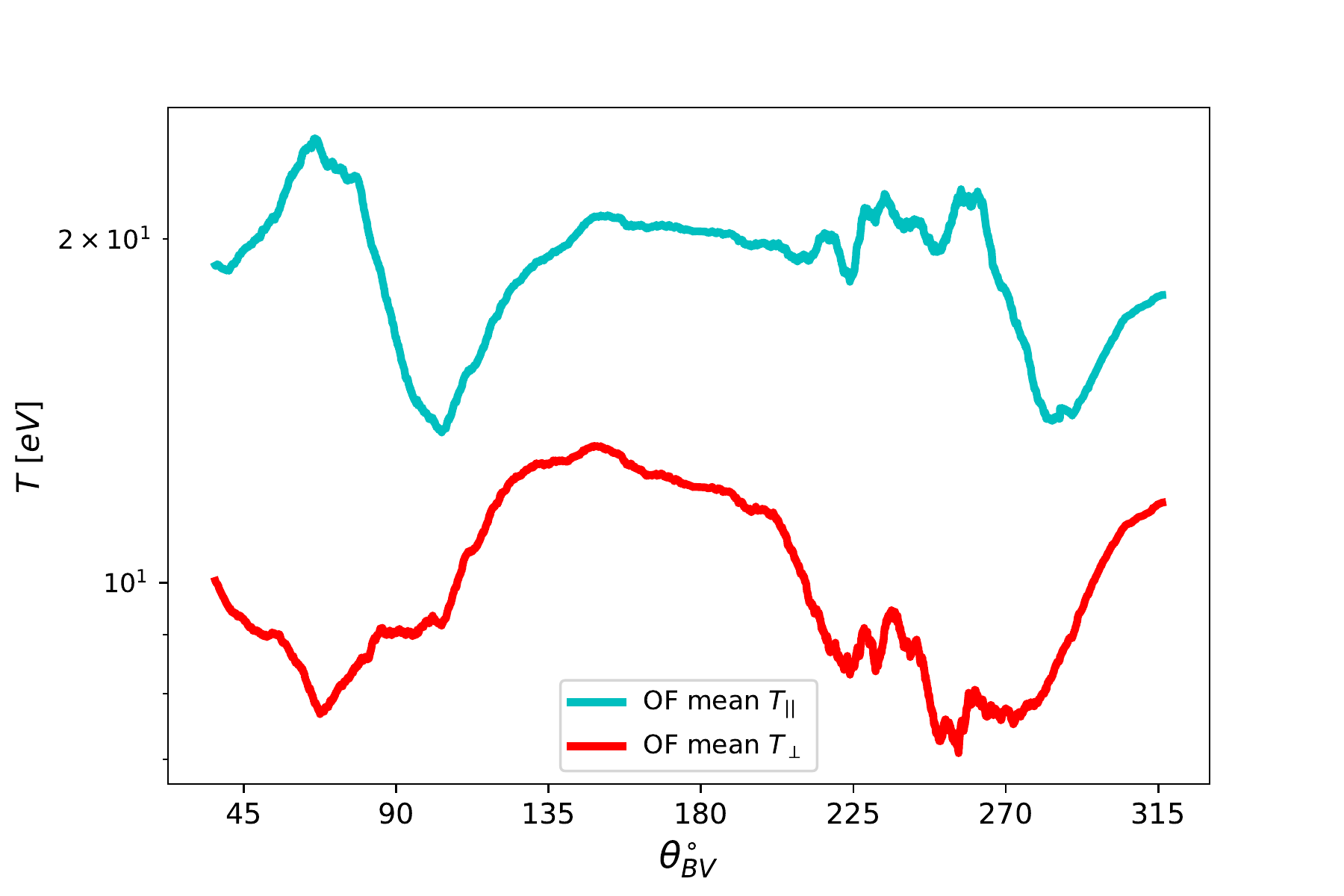}{0.5\textwidth}{(b)}
          }
\gridline{\fig{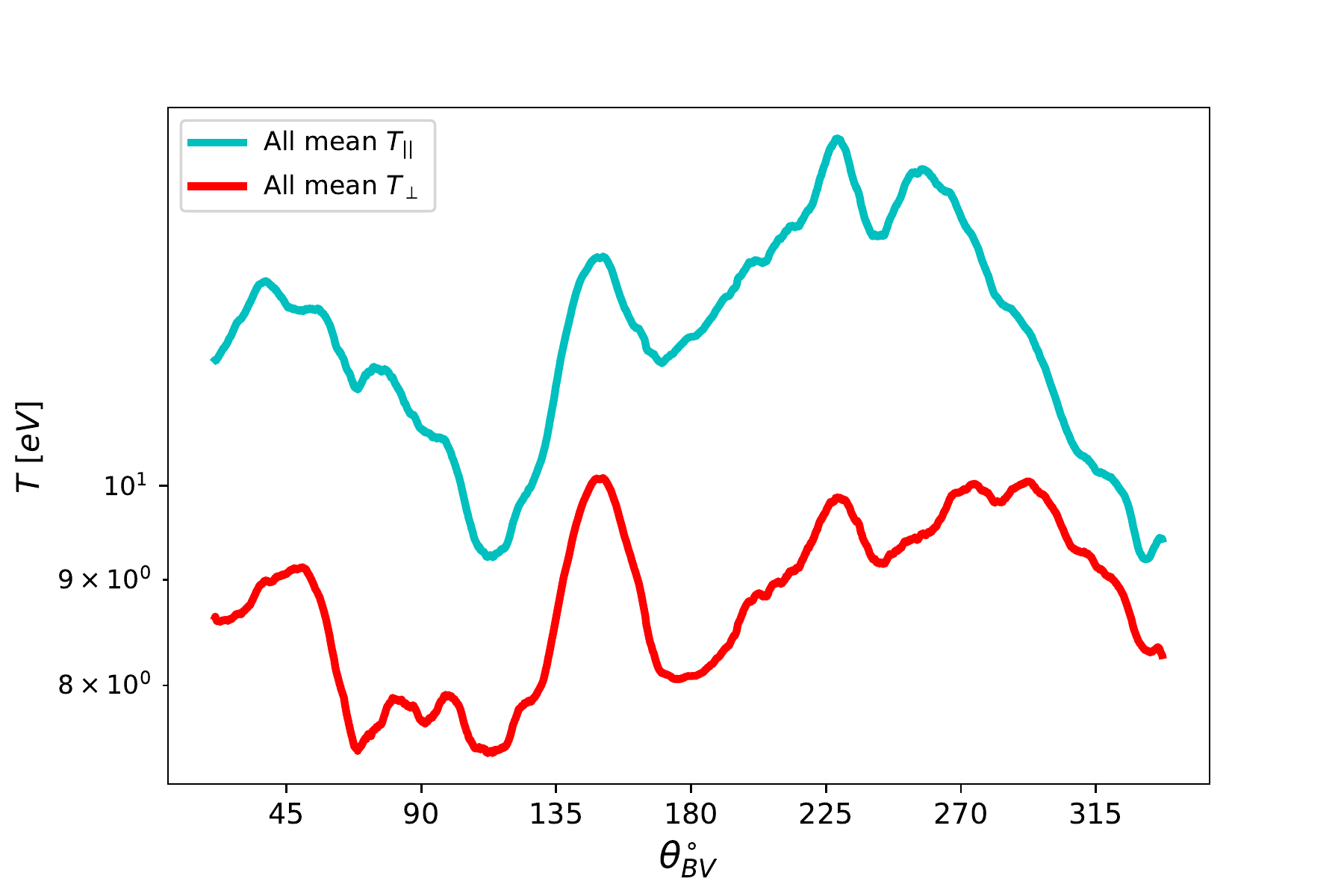}{0.5\textwidth}{(c)}
          \fig{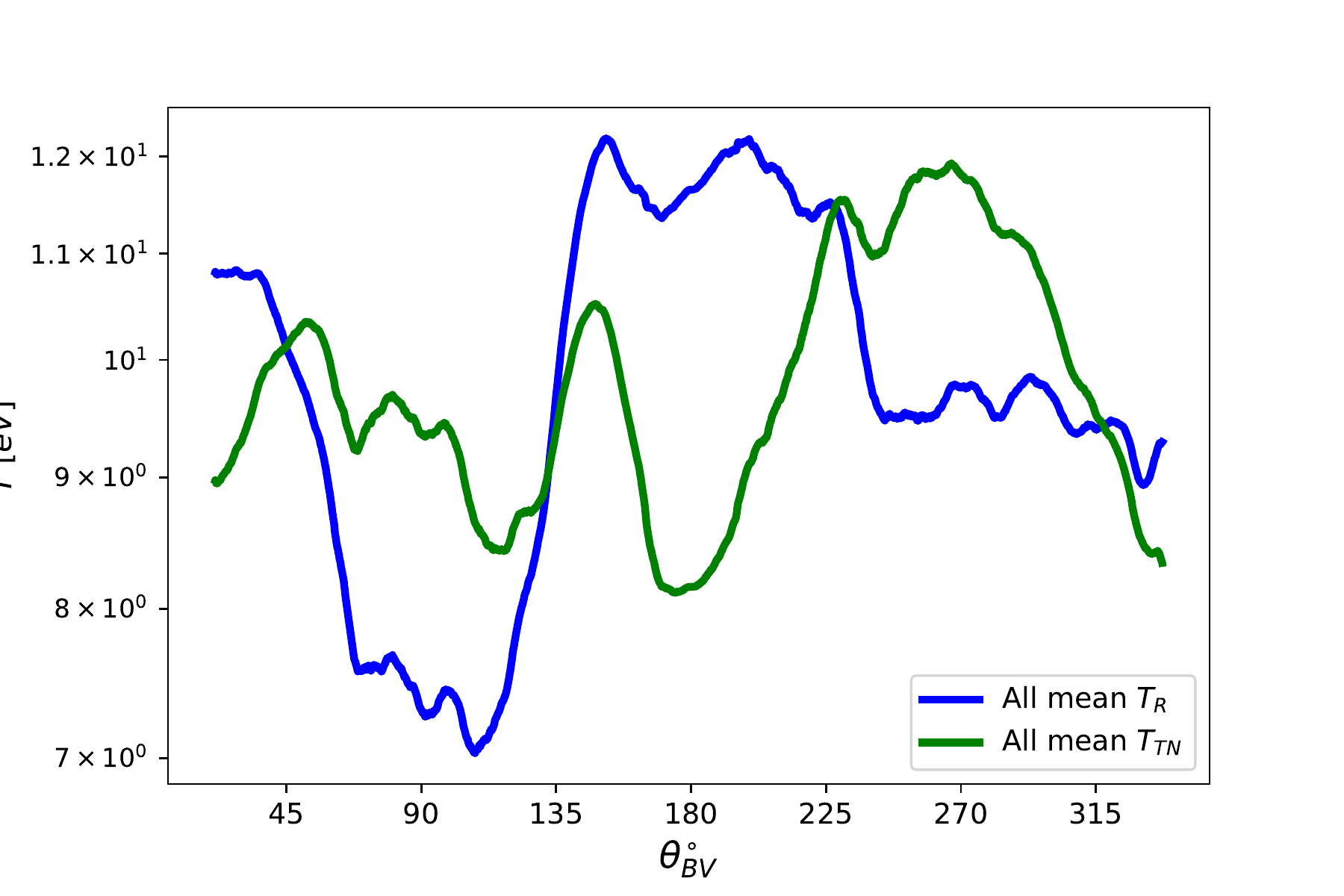}{0.5\textwidth}{(d)}
          }
\caption{Panels (a) to (c) show $T_\perp$ (red) and $T_{\parallel}$ (cyan) dependence on $\theta_{BV}$ in the solar wind for (a) mirror-mode, (b) oblique firehose unstable, and (c) all data. Panel (d) shows the equivalent dependency of $T_R$ (blue) and $T_{TN}$ (green) for all data. The result shown is a running mean over $\theta_{BV}$ after sorting the data points by $\theta_{BV}$. 
\label{fig:temp2}}
\end{figure*}

We explore the correlation between temperature anisotropy and $\theta_{BV}$ by investigating the mean values of $T_{\perp}$ and $T_{\parallel}$ as functions of $\theta_{BV}$ for the complete dataset and for each of the unstable categories. For this calculation, we first sort all $N_D$ data points in each of the data intervals by $\theta_{BV}$, where $N_D$ is the total number of data points in the interval. We then calculate a running mean over $\theta_{BV}$ for the separate parameters $T_{\perp}$ and $T_{\parallel}$ using a moving averaging window of length $0.04 N_D$. The results are plotted as Figure~\ref{fig:temp2}.

We show the running mean for each of $T_\perp$ and $T_\parallel$ for the mirror-mode instability, the oblique firehose instability, and for all data in panels (a) through (c). For comparison, we show a similar running mean of $T_R$ and $T_{TN}$ in panel (d), where $T_R$ is taken directly as the radial temperature from the $T_{RTN}$ dataset in the proton ground moments.  $T_{TN}$ is given by $(T_T+T_N)/2$, where $T_T$ is the tangential temperature and $T_N$ is the normal temperature.

\subsection{Lengthscale analysis}

\begin{figure*}
\gridline{\fig{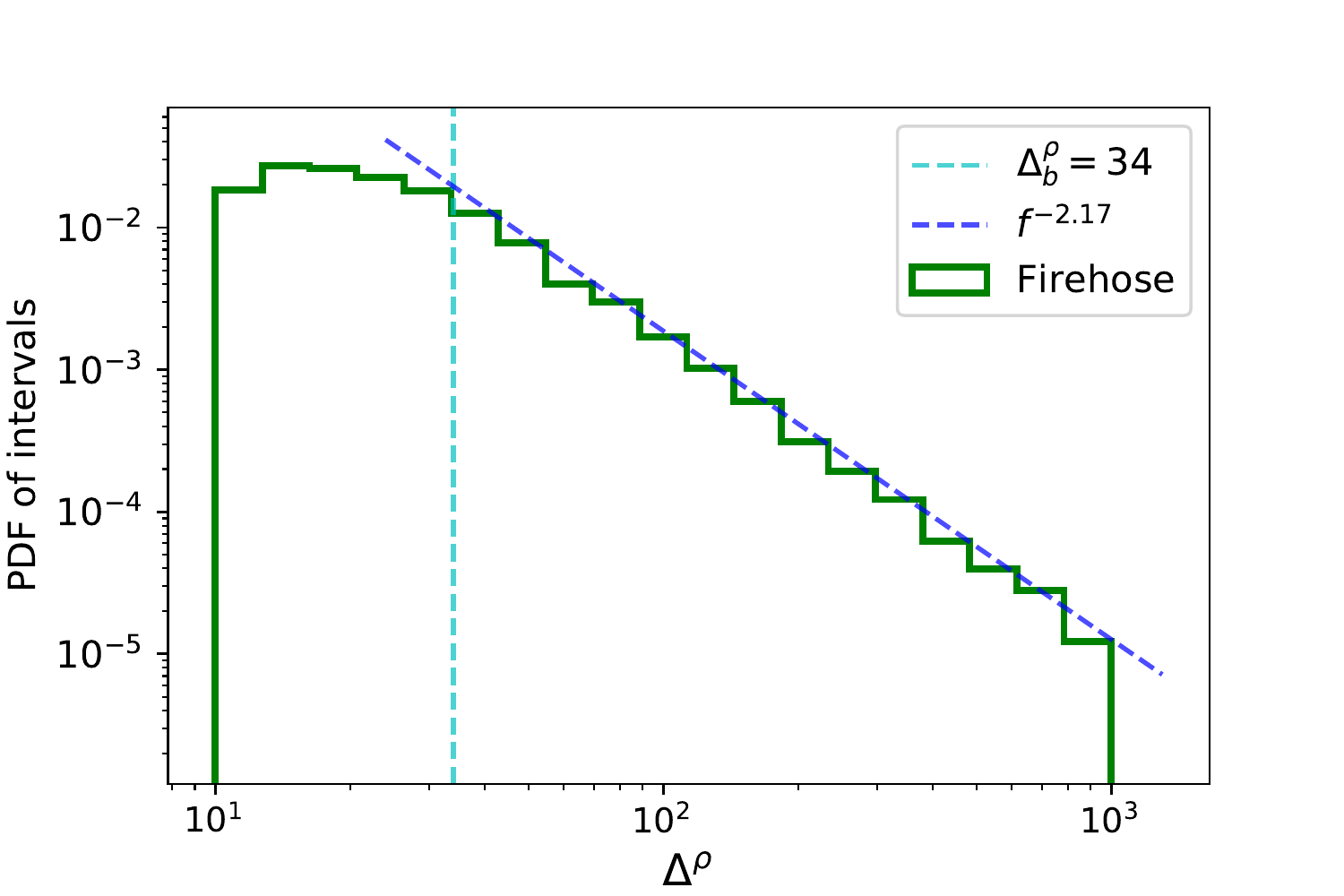}{0.5\textwidth}{(a)}
          \fig{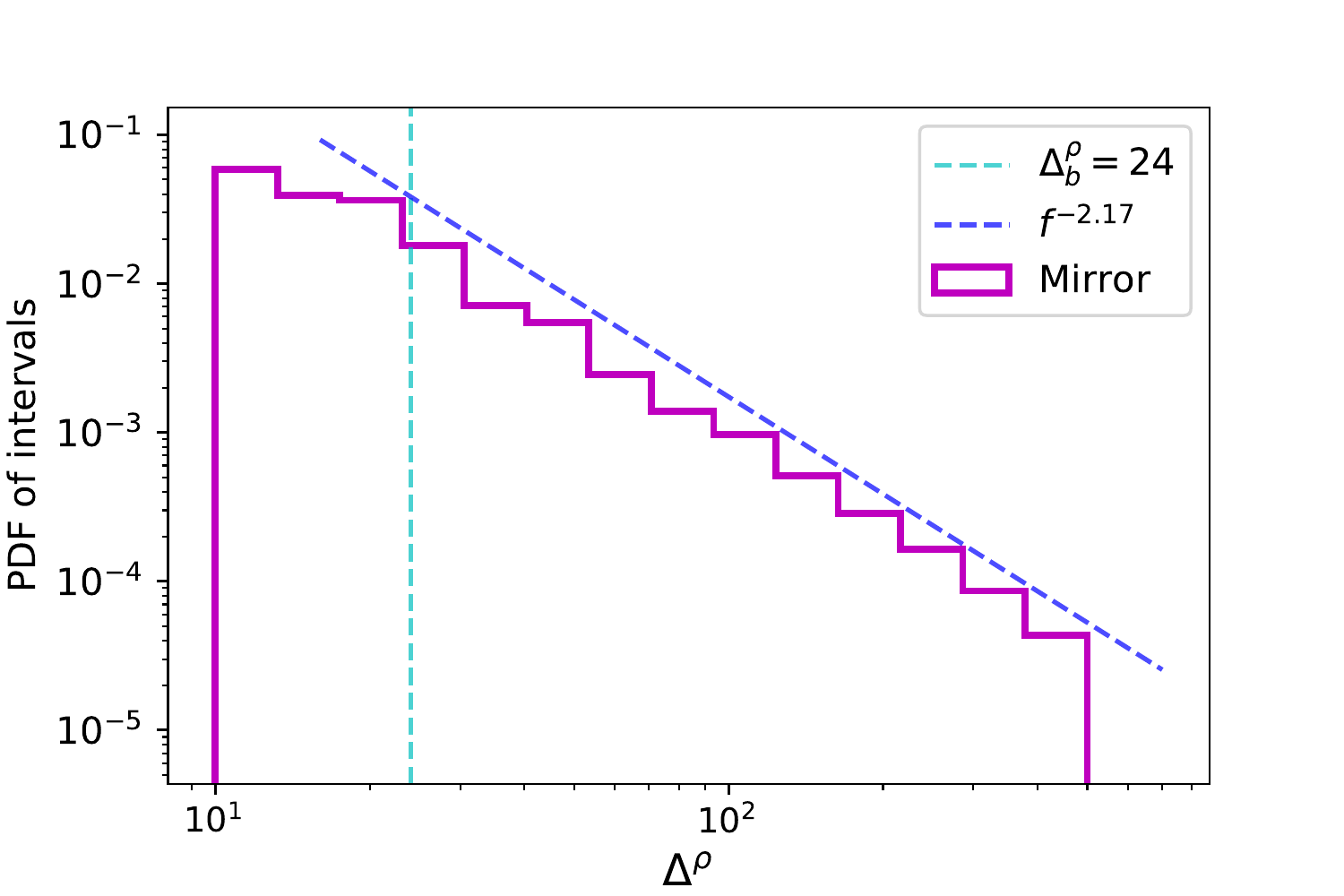}{0.5\textwidth}{(b)}
          }
\gridline{\fig{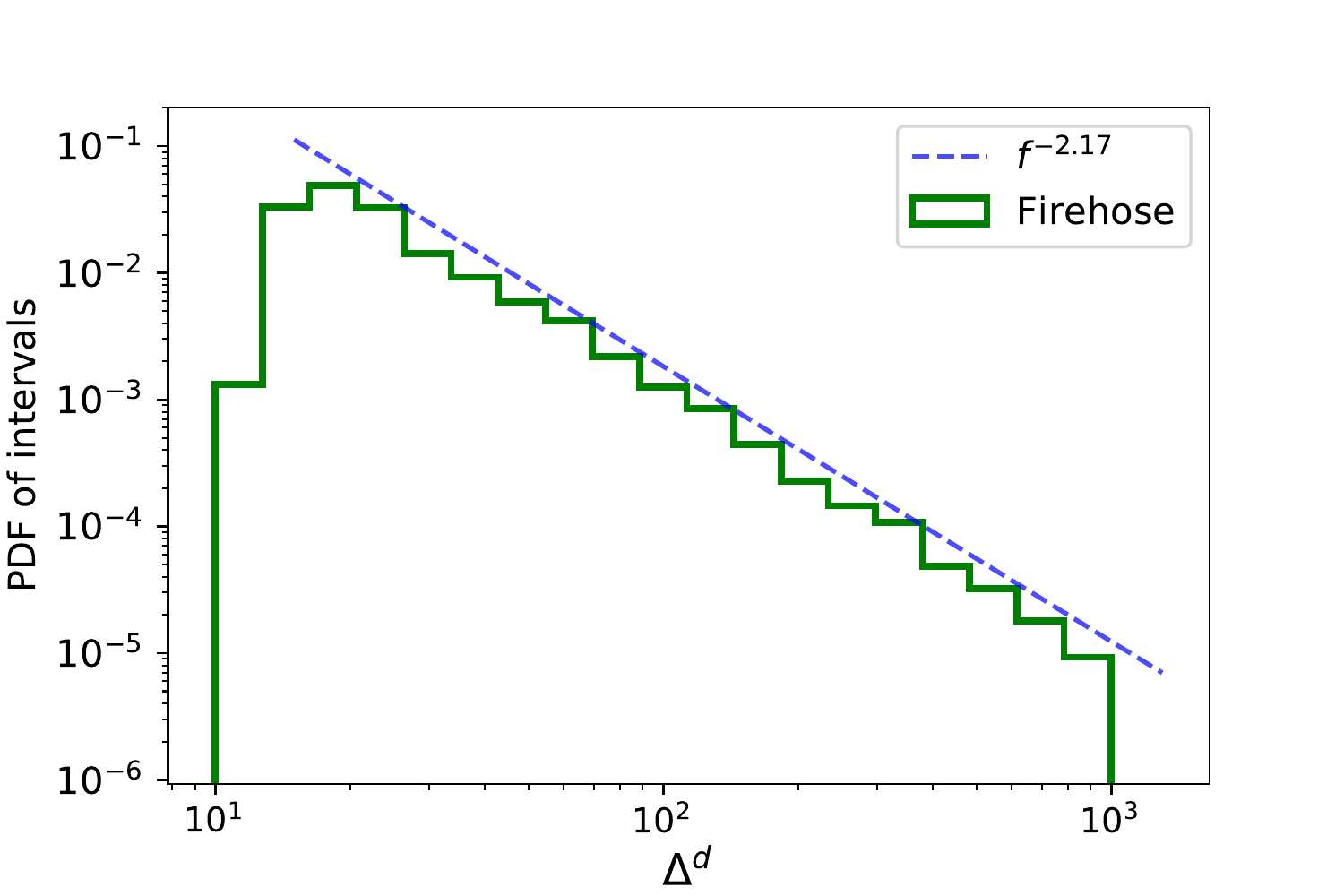}{0.5\textwidth}{(c)}
          \fig{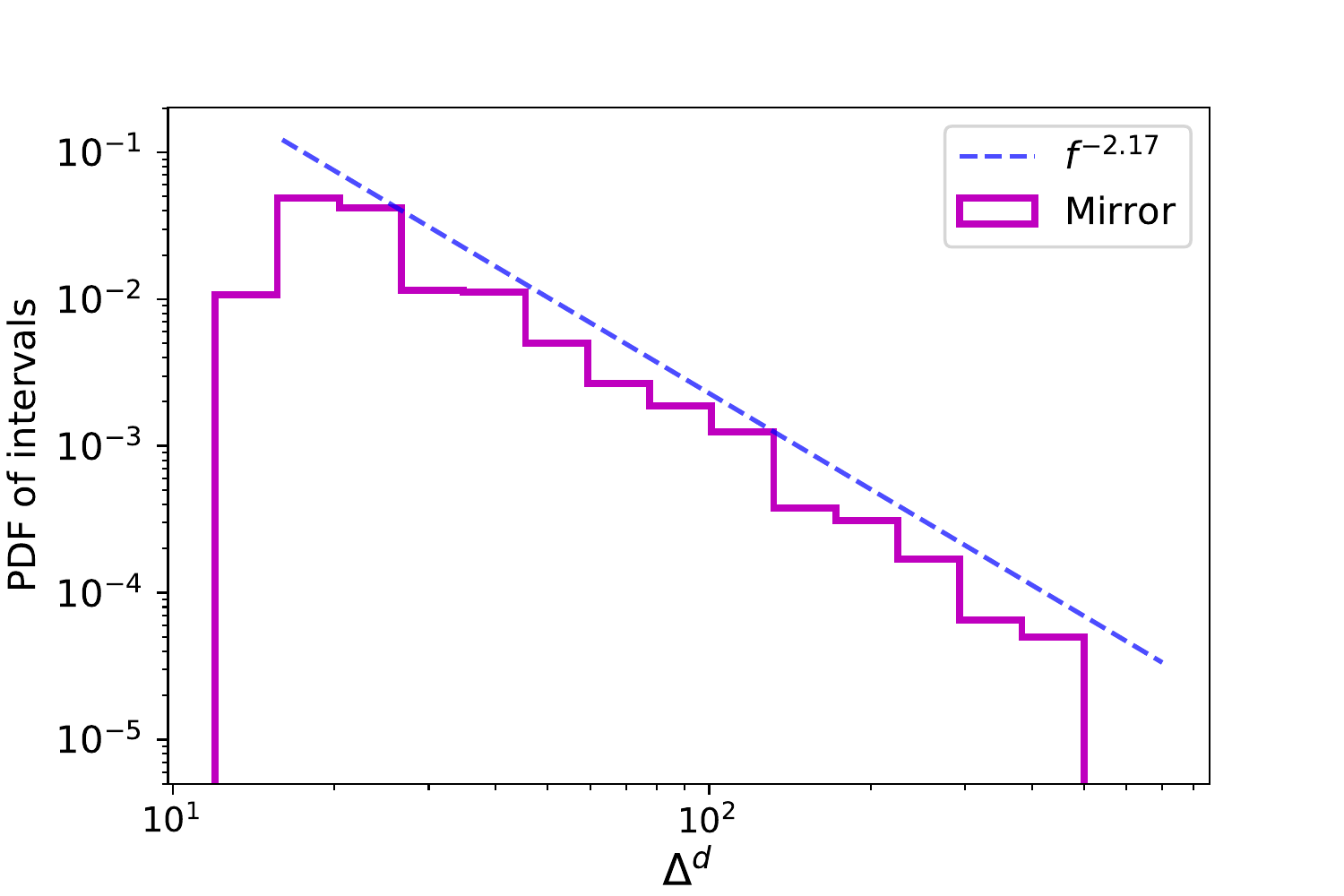}{0.5\textwidth}{(d)}
          }
\caption{PDF plots of persistence intervals in lengthscale for mirror-mode (magenta) and oblique firehose (green) instabilities in the solar wind. Panels (a) and (b) show the lengthscale in proton gyroradii ($\Delta^\rho$). Panels (c) and (d) show lengthscale in proton inertial lengths ($\Delta^d$). We indicate breakpoints at $\Delta_b^\rho$  with cyan vertical lines and power-law fits in dark blue, giving power indexes  of $\approx -2.17$ for both oblique firehose and mirror-mode instabilities.
\label{fig:persist}}
\end{figure*}

In Figure~\ref{fig:persist}, we show the PDFs of instability  persistence for both oblique firehose and mirror-mode instabilities measured in units  of the proton gyroradius  and in units of the inertial length. Panels (a) and (b) show a power-law relationship with a distinct break ($\Delta_b^\rho$) at $\sim 34\rho_p$ for the oblique firehose instability and at $\sim 24\rho_p$ for the  mirror-mode instability. At scales smaller than the break, the dependence of the instability persistence PDF on $\Delta^\rho$ exhibits a shallow gradient. At larger scales, the fitted power-law relationship is appreciably steeper and shows an exponent of $\approx -2.17$ for both the oblique firehose and the mirror-mode. This exponent is consistent with those found for $\Delta^d$ in panels (c) and (d), although here breakpoints are not readily identifiable. 

According to previous studies \citep{gary_theory_1993,pokhotelov_mirror_2004}, the maximum growth rate of the oblique firehose and mirror-mode instabilities occurs when $k_w\rho_{\mathrm p} \approx 0.5$, where $k_w$ is the wavenumber. 
The associated instability wavelength at maximum growth is then given by
    \begin{equation} \label{egro}
        \lambda_w =\frac{2\pi}{k_w}\approx \frac{2\pi \rho_{\mathrm p}}{0.5}.
    \end{equation}
Using $\Delta_b^{\rho}$ above, Eq.~(\ref{egro}) indicates that the space required for the instabilities to act is $\sim 2.7\lambda_w$ for the oblique firehose instability and $\sim 1.9\lambda_w$ for the mirror-mode instability.

\section{Discussion} \label{sec:discuss}

\subsection{Dependence on measurement cadence}

By contrast with most previous studies, we find a more extensive distribution of datapoints in the unstable regions of our $T_{\perp}/T_{\parallel}$-$\beta_{\parallel}$-plot in Figure~\ref{fig:bp}. This difference is likely due to the higher sampling cadence of PAS at $4\,\mathrm s$ in normal mode \citep{owen20} than, for example, that of the WIND SWE instrument. SWE has a cadence for its Faraday cup ion sensor of $92\,\mathrm s$ \citep{ogilvie_swe_1995,Maruca2013} and is the data source for many earlier studies \citep{bale09,chen_multi-species_2016,hellinger_solar_2006,maruca+2012,verscharen16}. PAS's higher cadence enables the instrument to sample characteristic features of the  solar wind without averaging over variations in the distribution at timescales greater than a few seconds \citep{verscharen_apparent_2011, nicolaou_impact_2019}. 

We demonstrate the impact of longer sampling times by averaging our dataset in Appendix~\ref{appaves}. In Figure~\ref{fig:averaging}, we show the consequent decrease in the proportions of datapoints in the mirror-mode and oblique firehose unstable regions with increasing measurement cadence. Although we observe in our data widespread distributions of data points in the parts of parameter space characterized as unstable according to Eq.~(\ref{ethresh}), the overall pattern of regulation by non-propagating instabilities ultimately remains consistent with the earlier work.  

\subsection{Interpretation of our angle analysis}

For our analysis, we divide the polar plot in Figure~\ref{fig:angles} into 8 segments of $45^\circ$ arc. We define the four segments between $315^\circ$ and $45^\circ$ and between $135^\circ$ and $225^\circ$ as `quasi-parallel' with respect to the flow direction of the solar wind. We define the other four segments as `quasi-perpendicular' with respect to the flow direction of the solar wind.

The oblique firehose instability is driven by $T_{\parallel}>T_{\perp}$ which, according to Figure~\ref{fig:angles} (b), most frequently corresponds to excess pressure in the direction quasi-perpendicular to the flow velocity. This direction is also approaching alignment with the Parker spiral angle. The mirror-mode instability is driven by $T_{\perp}>T_{\parallel}$ which, according to Figure~\ref{fig:angles} (c), most frequently also corresponds to excess pressure in the direction quasi-perpendicular to the flow velocity. This finding suggests that the expansion direction plays a crucial role for the generation of plasma conditions that drive both oblique firehose and mirror-mode instabilities. However, the large-scale double-adiabatic expansion alone according to the Chew--Goldberger--Low (CGL) prediction \citep{chew_boltzmann_1956, parker_dynamical_1958, matteini_evolution_2007, matteini_ion_2012} does not produce this observed correlation. In fact, the observed correlations are opposite to the expectations from CGL expansion alone, as shown in Appendix~\ref{appcgl}. Instead, we must invoke a non-CGL expansion of the ions, which is a known observational result \citep{marsch_temperature_2004,matteini_evolution_2007} and confirmed by the analysis of the two-fluid thermal energy equation \citep[e.g.][]{hellinger_proton_2013}.

A possible explanation for the angular dependency of the PDF for the oblique firehose and mirror-mode instabilities lies in the presence and variability of local, large-scale Alfv\'enic fluctuations. These fluctuations lead to a time variation in $\theta_{BV}$ at the location of the spacecraft \citep{damicis19}. 
Greater amplitudes of large-scale Alfv\'enic fluctuations typically coincide with increased perpendicular ion heating in the solar wind \citep{bruno_probability_2004,bruno_radial_2006}. This increased perpendicular heating often generates temperature anisotropy with $T_{\perp}>T_{\parallel}$ and thus favorable conditions for the excitation of mirror-mode instabilities \citep{matteini_parallel_2006,yoon_proton_2021}. As a consequence, we expect a statistically increased occurrence of mirror-mode unstable data points at times with large-amplitude Alfv\'enic fluctuations. These are more likely to be associated with $\theta_{BV}$ angles away from the direction of the Parker spiral than times without large-amplitude Alfv\'enic fluctuations \citep{bruno_probability_2004}.

In this interpretation, we associate oblique firehose unstable intervals with solar wind parcels with low amplitudes of Alfv\'enic fluctuations. We expect a stronger average alignment of the distribution of these oblique firehose unstable intervals with the average $\theta_{BV}$ of the solar wind. This is consistent with Figure~\ref{fig:angles} (a) and (b), which shows that the distribution of oblique firehose unstable intervals is mostly aligned  with the average direction of  $\vec B$, representing the Parker spiral angle \citep{parker_dynamical_1965} -- notwithstanding the asymmetry of sampled sector structures in our dataset. 

Figure~\ref{fig:temp2} (a) through (c) demonstrate that the measured solar wind on average exhibits $T_\parallel > T_\perp$, so that conditions favorable for the excitation of the mirror-mode instability are the exception. However, panel (c) shows that for all data $T_\parallel$ and  $T_\perp$ converge when $\vec B$ and $\vec V$ approach alignment or anti-alignment, which is consistent with the statistical distribution of the unstable datapoints we observe in Figure~\ref{fig:angles2}. However, the sector asymmetry of our dataset makes this convergence stronger in the case of $\theta_{BV} \sim 360^\circ$.

The distributions of the unstable data as a function of $\theta_{BV}$ shown in Figure~\ref{fig:angles} (b) and (c), together with the required temperature anisotropy to drive each of the instabilities, suggest that the mirror-mode and oblique firehose instabilities act predominantly when $T_{TN} > T_R$, where for the mirror-mode $T_{TN} \sim T_\perp$ and for the oblique firehose $T_{TN} \sim T_\parallel$. The  mean values of the dataset show that, on average,  $T_R/T_{TN}$ = 0.985 (all data), 0.447 (oblique firehose unstable data points), and 0.572 (mirror-mode unstable data points). Figure~\ref{fig:temp2} (d) shows the $\theta_{BV}$-dependence in the variability of $T_R$ and $T_{TN}$ which derives from the general condition that $T_\parallel > T_\perp$ on average as shown in Figure~\ref{fig:temp2} (c). The angular dependency of the observed oblique firehose unstable data is consistent with this variability. Whereas the distribution of mirror-mode unstable data peaks at the values of $\theta_{BV}$ where the plots of $T_R$ and $T_{TN}$ intersect. We observe that these points of intersection are the limits to the values of $\theta_{BV}$ where both $T_{TN} \sim T\perp$ and $T_{TN} \geq T_R$. 

\subsection{Interpretation of our lengthscale analysis}

In Figure~\ref{fig:persist}, we find that our PDF of $\Delta^\rho$ steepens appreciably at the breakpoint $\Delta^\rho_b$. The breakpoint is more clearly defined by lengthscales normalized in units of $\rho_p$  than in units of $d_p$, which is expected since these instabilities grow on scales associated with the gyroscale rather than the inertial length \citep{howes_gyrokinetic_2011, matthaeus_nonlinear_2014}.
We interpret the shallower PDF at $\Delta ^{\rho}<\Delta_b^\rho$ as an indication that, in spatial intervals shorter than $\Delta_b^{\rho}$, the instabilities are less efficient in reducing the temperature anisotropy to stable values than in intervals longer than $\Delta_b^\rho$. The existence of this breakpoint and the transition into a steeper slope at $\Delta^\rho>\Delta_b^\rho$ is consistent with our conjecture that the efficiency of oblique firehose and mirror-mode instabilities is scale-dependent.
In this interpretation, $\Delta_b^\rho$ represents  the minimum length of plasma intervals with unstable parameters for which the instabilities efficiently modify the plasma into a stable state. 
We interpret the difference in $\Delta_b^\rho$ between oblique firehose and mirror-mode instabilities as an indication that these instabilities set different requirements on the homogeneity of the unstable plasma volumes. We observe that the power-law index beyond the breakpoint is consistent for both categories of unstable data and independent of our lengthscale normalization. This universality suggests that the power laws themselves are  representative of the underlying distribution of conditions that drive the analyzed instabilities in the solar wind.

We find that $\Delta_b^\rho$ corresponds to approximately 2 to 3 wavelengths of the unstable mode at typical maximum growth rates. This result suggests that the conditions needed for instabilities to act efficiently are bounded by spatial scales $\mathcal{O}(10^3\, \mathrm{km})$ that are very short relative to the correlation length of solar wind turbulence which has been measured as $\mathcal{O}(10^6\, \mathrm{km})$ \citep{matthaeus_measurement_1982, matthaeus_spatial_2005}. However, the evaluation of the influence of the turbulent cascade on linear processes requires a scale-dependent comparison of nonlinear and linear timescales  \citep{matthaeus_nonlinear_2014}. While outside the scope of this study, it would be worthwhile to compare the scale-dependent eddy turnover times of the plasma turbulence at the scales of the unstable intervals. Such a comparison would allow the assessment of the timescales that potentially create and destroy the conditions required for instabilities to act \citep{klein_applying_2017,klein_majority_2018,qudsi_observations_2020}.

In our study, we assume Taylor's hypothesis to link temporal variations in the measurements with spatial variations in the solar wind. A single-spacecraft measurement is unable to disentangle temporal and spatial variations. Simulations show a temporal latency in the onset of the oblique firehose instability \citep{lopez_mixing_2022}, which complicates the interpretation of the spatial latency discussed in this work.
The average temporal persistence of datapoints in the unstable regions of our $T_{\perp}/T_{\parallel}$-$\beta_{\parallel}$-plot  is  $10.33\,\mathrm{s}$ for the mirror-mode and $11.49\,\mathrm{s}$ for the  oblique firehose instability. From these numbers, we infer that a sampling cadence of less than $\sim 10\,\mathrm{s}$ is needed to observe the full lengthscale distribution of unstable regions in the solar wind in our dataset. This cadence is equivalent to a spatial scale of $\sim 80\rho_{\mathrm p} $ convected over the spacecraft, given the average values for $V$ and $\rho_{\mathrm p}$ derived from our data. 

\subsection{Limitations of our analysis}

Our analysis is necessarily limited by the statistics and quality of the  dataset. We use data from the cruise phase of the Solar Orbiter mission \citep{zouganelis2020}. Our period of data collection coincides with relatively quiet solar wind conditions, and the dataset contains mostly slow solar wind observations.  Instrumental effects on our observations have been carefully evaluated in consultation with the SWA and MAG teams. Our rigorous application of the available quality filters to the data and our exchanges with the instrument teams increase the reliability of our analysis. Our result in Figure~\ref{fig:bp} depends on the details of the method used to define the instability thresholds \citep[see also][]{isenberg_self-consistent_2013}. Our method is however consistent with previous studies concerning the specific instabilities we consider \citep{hellinger_solar_2006, bale09}. Our Figure~\ref{fig:temp2} differs from the more straightforward $T_{\perp}$-dependence presented by \citet{damicis19} who find a positive correlation between $\theta_{BV}$ and $T_\perp$ in Alfv\'enic fast solar wind. However, this result does not contradict our analysis, given that we mostly observe slow solar wind in our dataset.

\section{Conclusions} \label{sec:concs}

We perform a statistical analysis of a large Solar Orbiter dataset ($\sim 10^6$ data points) to investigate the conditions necessary in the solar wind for the  oblique firehose and mirror-mode instabilities to reduce temperature anisotropies. Our motivation is to use the newly available high-resolution data from the Solar Orbiter mission to explore energy transfer processes at small scales.

In our $T_{\perp}/T_{\parallel}$-$\beta_{\parallel}$-plot, we find that, while the investigated instabilities largely limit the plasma anisotropy, a significant number of data points ($\sim 3\times10^4$ for the oblique firehose and $\sim 4\times10^3$ for the mirror-mode) lie in the unstable regions of parameter space. We interpret these as transient features whose full extent is revealed by the short measurement time ($1\,\mathrm{s}$) of the SWA instrument's PAS \citep{owen20}. 

We explore the dependency of the distribution of  oblique firehose and mirror-mode unstable solar wind intervals  on $\theta_{BV}$ and find that the mirror-mode instability predominantly occurs when $\theta_{BV}\approx 0^{\circ} \pm 45^{\circ}$ or $\theta_{BV}\approx 180^{\circ} \pm 45^{\circ}$ and hence when $\vec B$ is close to the radial (or anti-radial) direction. By contrast, the peak in the PDF  of the oblique firehose instability occurs when $\theta_{BV}\approx 75^{\circ}$  or $\theta_{BV}\approx 255^{\circ}$ and hence when $\vec B$ is close to a direction perpendicular to $\vec V$. This result suggests a predominant elevation of the temperature $T_{TN}$ perpendicular to the radial direction relative to the temperature $T_R$ in the radial direction in the unstable intervals, which is inconsistent with the predictions from the CGL double-adiabatic expansion of the solar wind alone. We interpret this dependency of the mirror-mode (oblique firehose) unstable plasma intervals on $\theta_{BV}$ as due to the presence (absence) of perpendicular ion heating from local, large-scale Alfv\'enic fluctuations.

In our analysis of $\theta_{BV}$, we do not confine consideration to Alfv\'enic wind intervals \citep{louarn21, damicis19, woodham21} but instead concentrate on the relationship between $\theta_{BV}$ and the specific proton temperature anisotropy that drives the  oblique firehose and  mirror-mode instabilities. This allows us to investigate the conditions  needed for instabilities to act efficiently on the solar wind  plasma and possible explanations for the occurrence of these conditions. However, our analysis is necessarily constrained by the statistical $\theta_{BV}$ distribution in our dataset, which we expect to become less asymmetrical as more data are included while the Solar Orbiter mission continues.

We also measure the persistence of unstable solar wind intervals. The oblique firehose and mirror-mode instabilities require intervals of a size greater than about $34\rho_p$ and $24\rho_p$, respectively, in order to regulate the temperature anisotropy efficiently. These lengthscales are more clearly defined in units of $\rho_p$ rather than in units of $d_{p}$. The minimum space to develop and regulate anisotropy corresponds to approximately 2 to 3 typical wavelengths of the unstable mode at maximum growth rate. 

Our work highlights the intricate connections between expansion effects, turbulence, and kinetic micro-instabilities in the solar wind. Numerical simulations show that a combination of plasma expansion and strong 2D turbulence can drive both oblique firehose and mirror-mode instabilities  \citep[][]{hellinger_plasma_2015,hellinger_mirror_2017}. In addition, the spread of data in the  $T_{\perp}/T_{\parallel}$-$\beta_{\parallel}$ parameter space is increased by pronounced small-scale intermittency in strong turbulence \citep{servidio_proton_2014}. The combination of in-situ instruments on the Solar Orbiter mission allows us to study particle distributions with a very high time resolution, which helps to gain fresh insight into the underlying processes \citep{adhikari21,damicis_2021,louarn21, nicolaou21orbiter,owen_solar_2021}. We expect high-cadence in-situ observations in combination with kinetic simulations of the expanding solar wind \citep{dong_evolution_2014,hellinger_plasma_2015,franci_solar_2015} to deliver further insights into this interplay in the future.

%\section{Conclusions} \label{sec:concs}

\begin{acknowledgments}

We appreciate very helpful discussions with the MAG and PAS instrument teams.
We thank Thomas Keel, Kris Klein, and Rob Wicks for very useful discussions and advice, and Jonathan Niehof for very helpful personal correspondence on the application of SpacePy's pycdf toolkit.
S.~O.~is supported by the Natural Environment Research Council (NERC) grant  NE/S007229/1. D.~V.~is supported by the Science and Technology Facilities Council (STFC) Ernest Rutherford Fellowship ST/P003826/1. D.~V.~and C.~J.~O. are supported by STFC Consolidated Grants  ST/S000240/1 and ST/W001004/1. C.~H.~K.~C.~is supported by UKRI Future Leaders Fellowship MR/W007657/1, STFC Ernest Rutherford Fellowship ST/N003748/2 and STFC Consolidated Grant ST/T00018X/1. P.~A.~I.~is supported by NASA grant 80NSSC18K1215 and by National Science Foundation (NSF) grant AGS2005982.

Solar Orbiter is a space mission of international collaboration
between ESA and NASA, operated by ESA. Solar Orbiter Solar Wind
Analyser (SWA) data are derived from scientific sensors which have been
designed and created, and are operated under funding provided in
numerous contracts from the UK Space Agency (UKSA), STFC, the Agenzia Spaziale Italiana
(ASI), the Centre National d’Etudes Spatiales (CNES), the Centre
National de la Recherche Scientifique (CNRS), the Czech
contribution to the ESA PRODEX programme and NASA. Solar Orbiter SWA
work at UCL/MSSL is currently funded under STFC grants ST/T001356/1 and
ST/S000240/1. The Solar Orbiter magnetometer was funded by UKSA grant ST/T001062/1.

\end{acknowledgments}



\appendix

\section{Averaging effect} \label{appaves}

\begin{figure*}
\gridline{\fig{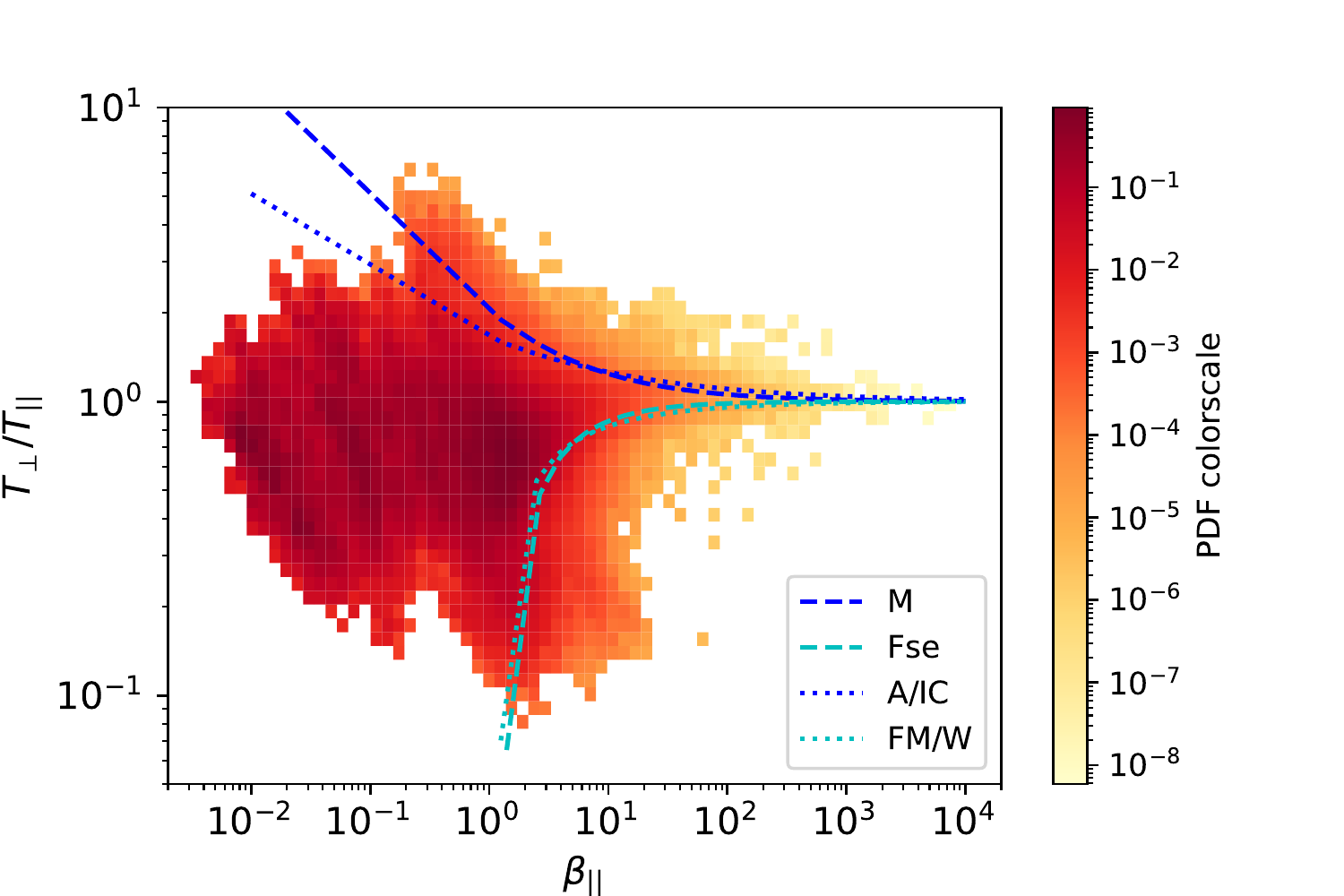}{0.5\textwidth}{(a)}
          \fig{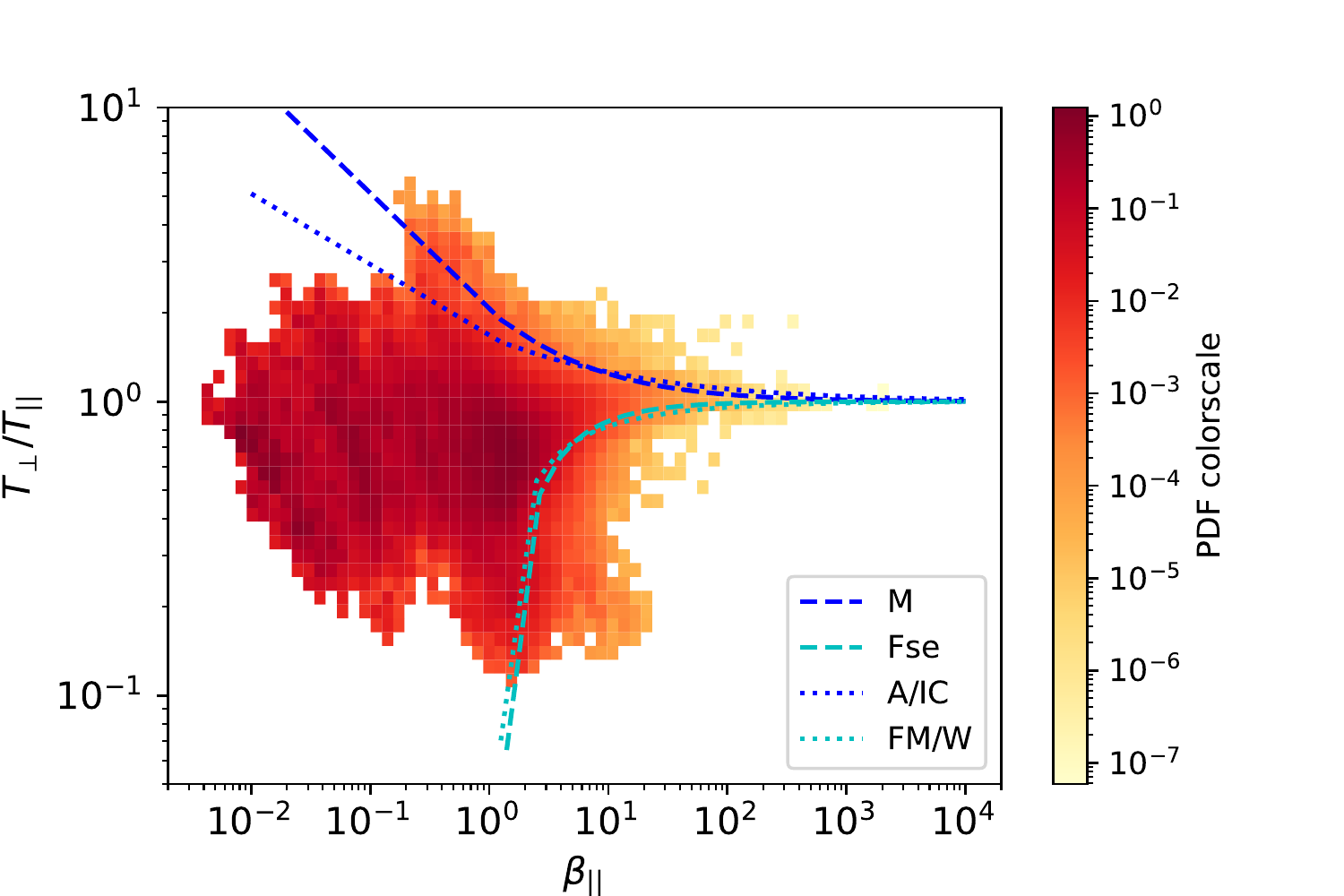}{0.5\textwidth}{(b)}
          }
\gridline{\fig{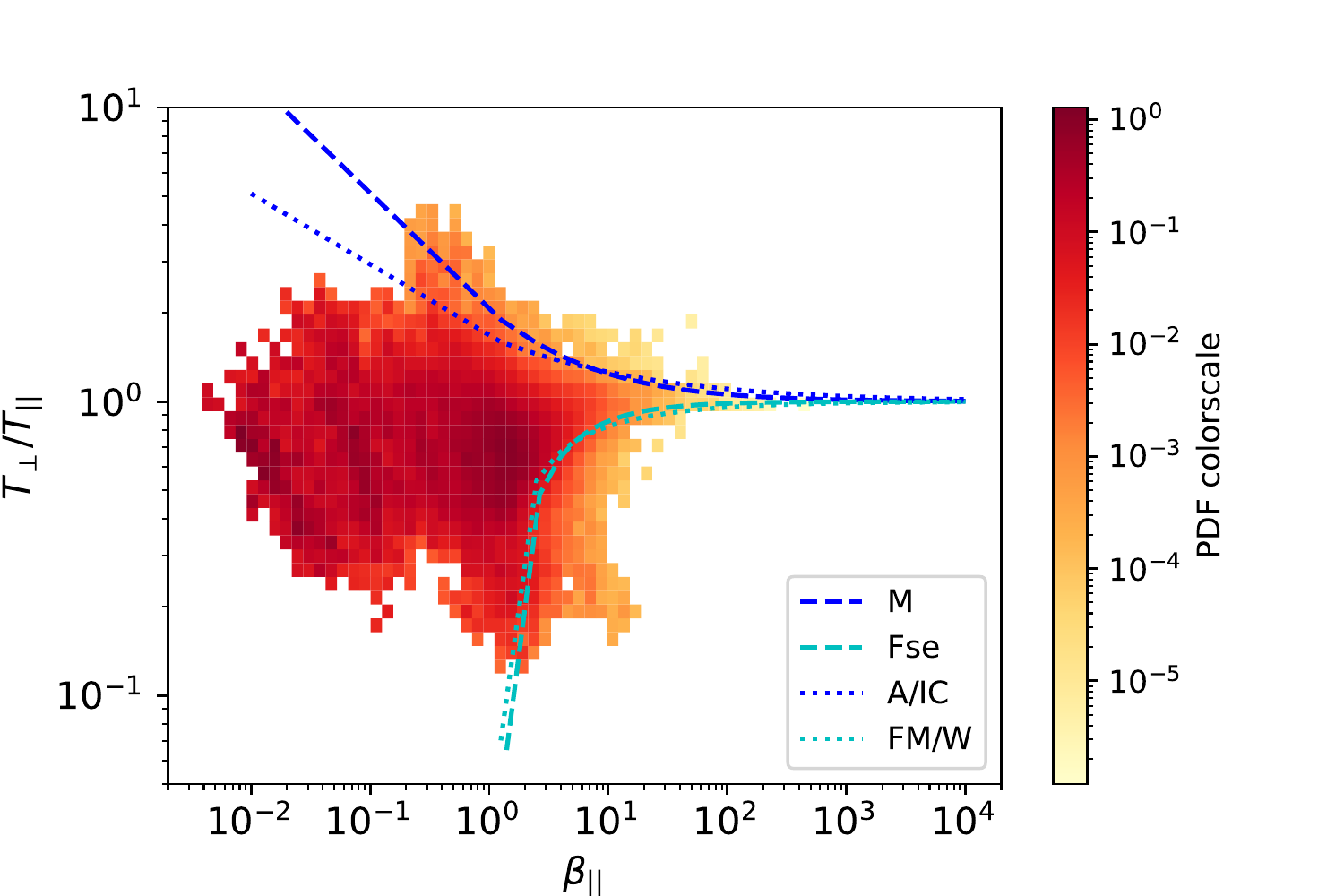}{0.5\textwidth}{(c)}
          \fig{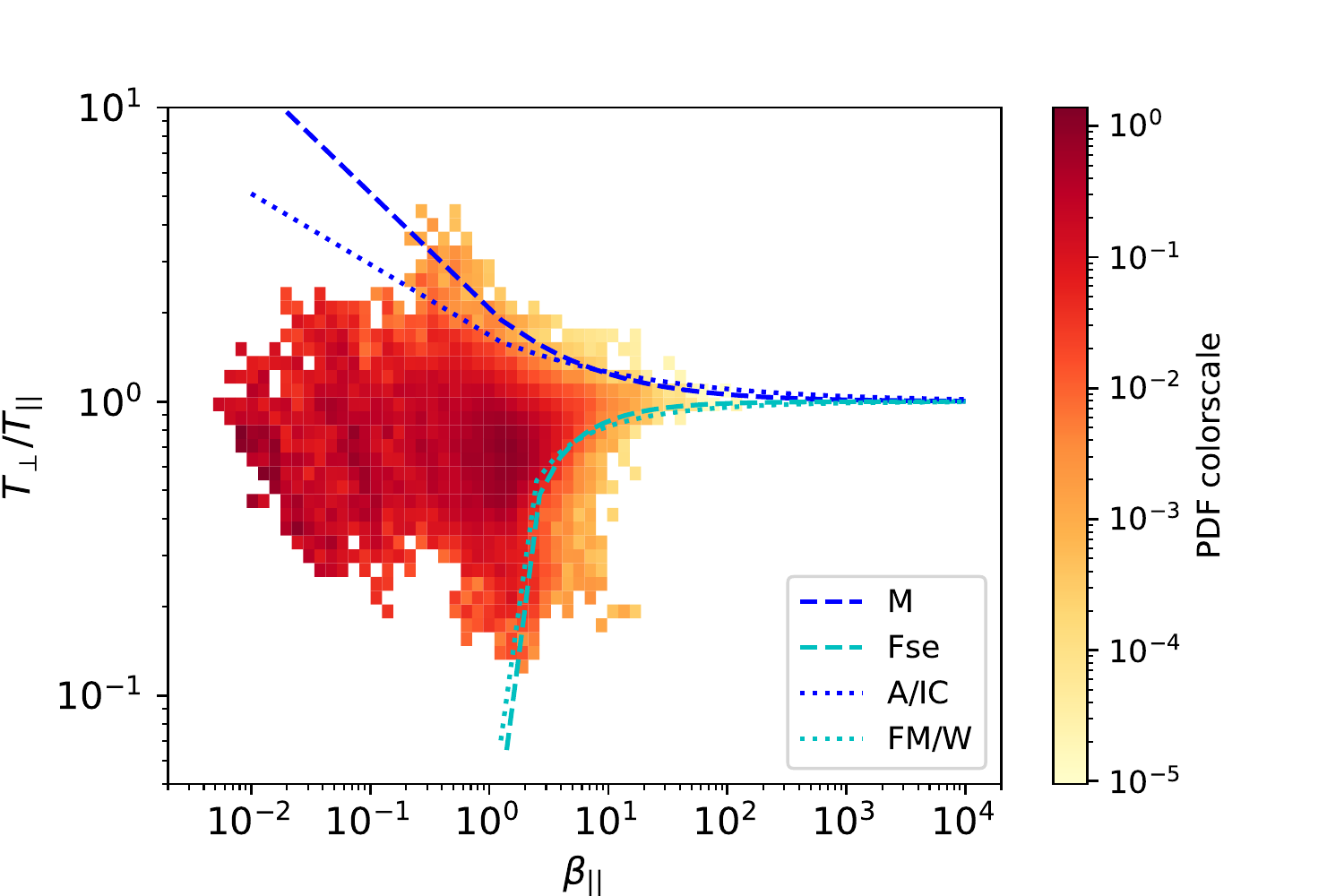}{0.5\textwidth}{(d)}
          }
\caption{$T_\perp/T_{\parallel}$-$\beta_{\parallel}$-plots for averaged solar wind parameters with different averaging times. We overplot the instability thresholds according to Eq.~(\ref{ethresh}) for $\gamma_m  = 10^{-2} \Omega_p$. The color-coding indicates the  PDF of datapoints in each bin. The different panels represent averaging of data over subsequent sampling steps to simulate instruments with lower measurement cadences: (a) $4\,\mathrm{s}$ (original dataset), (b) $12\,\mathrm{s}$, (c) $48\,\mathrm{s}$, and (d) $92\,\mathrm{s}$. 
\label{fig:averaging}}
\end{figure*}

Our analysis in Figure~\ref{fig:persist} suggests that plasma instruments with low measurement cadence detect a lower proportion of unstable intervals than actually exist. In general, all instruments miss unstable intervals with a duration in the spacecraft frame comparable to the measurement cadence or shorter.

We simulate different measurement cadences by averaging our data over a successive number of sampling intervals. In Figure~\ref{fig:averaging}, we show the results averaged over 12, 48, and 92\,s, compared with the base dataset at 4-s cadence. The proportional share of datapoints in the regions of parameter space unstable to oblique firehose and mirror-mode instabilities increases with increasing cadence.
At 4~s cadence, 3.12\% of the data are oblique firehose unstable and 0.46\% of the data are mirror-mode unstable. At 92\,s, which corresponds approximately to the sampling cadence of the SWE instrument onboard WIND, these numbers decrease to 2.45\% for the oblique firehose instability and 0.31\% for the mirror-mode instability. Moreover, the number of data points with extreme $\beta_{\parallel}$-values decreases significantly.

\section{CGL Analysis} \label{appcgl}

The double-adiabatic expansion according to the CGL theory is often considered an important contributor to the impact of expansion on the temperature anisotropy of the solar wind  \citep{matteini_evolution_2007,verscharen16}. In this Appendix, we evaluate the consistency of the CGL approach with our observations.

We start by assuming that the solar-wind response to the expansion is consistent with the CGL equations \citep{chew_boltzmann_1956}:
\begin{equation}\label{cgl1}
    \frac{d}{dt}\left(\frac{T_\perp}{B}\right) = 0
\end{equation}
and
\begin{equation}\label{cgl2}
    \frac{d}{dt}\left(\frac{T_{\parallel}B^2}{n_p^2}\right) = 0.
\end{equation}
When $\vec B$ is purely in the radial direction, $B \propto r^{-2}$ due to $\nabla \cdot B=0$. Likewise, when $\vec B$ is purely in the tangential direction, $B \propto r^{-1}$ in a spherically symmetric configuration \citep{matteini_evolution_2007,matteini_ion_2012,hellinger_plasma_2015}.
From Eqs.~(\ref{cgl1}) and (\ref{cgl2}), we obtain
\begin{equation}
    \frac{T_\perp}{T_{\parallel}} \propto \frac{B^3}{n_p^2}.
\end{equation}
Hence, under the assumption that $\vec B$ is purely radial, we find
\begin{equation}\label{ani_rad}
    \frac{T_\perp}{T_{\parallel}} \propto r^{-2}.
\end{equation}
Under the assumption that  $\vec B$ is purely tangential, we find
\begin{equation}\label{ani_iso}
    \frac{T_\perp}{T_{\parallel}} \propto r.
\end{equation}
According to Eqs.~(\ref{ani_rad}) and (\ref{ani_iso}), CGL double-adiabatic expansion in a smooth magnetic field predicts that $T_\perp/T_\parallel$ drops more quickly with $r$ when $\vec B$ is in the radial direction than when $\vec B$ is in the tangential direction. 

Assuming that the solar wind magnetic field obeys Parker's model, in which the Parker angle is a monotonously increasing  function of $r$ \citep{parker_dynamical_1965}, we expect conditions favorable for the oblique firehose instability especially when the field is quasi-radial. However, Figure~\ref{fig:angles} reveals the opposite behavior and even a higher occurrence of mirror-mode unstable intervals in the quasi-radial field geometry.
This finding suggests that CGL double-adiabatic expansion alone does not provide a consistent explanation for the angular dependency result shown in Figure~\ref{fig:angles}.  The observed distribution of these instabilities contradicts the CGL prediction, suggesting that other processes must dominate the observed occurrence rates of unstable intervals.

%\clearpage

\bibliography{paper1}

\begin{thebibliography}{}
\expandafter\ifx\csname natexlab\endcsname\relax\def\natexlab#1{#1}\fi
\providecommand{\url}[1]{\href{#1}{#1}}
\providecommand{\dodoi}[1]{doi:~\href{http://doi.org/#1}{\nolinkurl{#1}}}
\providecommand{\doeprint}[1]{\href{http://ascl.net/#1}{\nolinkurl{http://ascl.net/#1}}}
\providecommand{\doarXiv}[1]{\href{https://arxiv.org/abs/#1}{\nolinkurl{https://arxiv.org/abs/#1}}}

\bibitem[{Adhikari {et~al.}(2021)Adhikari, Zank, Zhao, Telloni, Horbury,
  O'Brien, Evans, Angelini, Owen, Louarn, \& Fedorov}]{adhikari21}
Adhikari, L., Zank, G.~P., Zhao, L.-L., {et~al.} 2021, Astronomy \&
  Astrophysics, \dodoi{10.1051/0004-6361/202140672}

\bibitem[{Alexandrova {et~al.}(2013)Alexandrova, Chen, Sorriso-Valvo, Horbury,
  \& Bale}]{alexandrova_solar_2013}
Alexandrova, O., Chen, C. H.~K., Sorriso-Valvo, L., Horbury, T.~S., \& Bale,
  S.~D. 2013, Space Science Reviews, 178, 101,
  \dodoi{10.1007/s11214-013-0004-8}

\bibitem[{{Bale} {et~al.}(2009){Bale}, {Kasper}, {Howes}, {Quataert}, {Salem},
  \& {Sundkvist}}]{bale09}
{Bale}, S.~D., {Kasper}, J.~C., {Howes}, G.~G., {et~al.} 2009, \prl, 103,
  211101, \dodoi{10.1103/PhysRevLett.103.211101}

\bibitem[{Bruno {et~al.}(2006)Bruno, Bavassano, D’amicis, Carbone,
  Sorriso-Valvo, \& Pietropaolo}]{bruno_radial_2006}
Bruno, R., Bavassano, B., D’amicis, R., {et~al.} 2006, Space Science Reviews,
  122, 321, \dodoi{10.1007/s11214-006-5232-8}

\bibitem[{Bruno \& Carbone(2013)}]{bruno_solar_2013}
Bruno, R., \& Carbone, V. 2013, Living Reviews in Solar Physics, 10,
  \dodoi{10.12942/lrsp-2013-2}

\bibitem[{Bruno {et~al.}(2004)Bruno, Carbone, Primavera, Malara, Sorriso-Valvo,
  Bavassano, \& Veltri}]{bruno_probability_2004}
Bruno, R., Carbone, V., Primavera, L., {et~al.} 2004, Annales Geophysicae, 22,
  3751, \dodoi{10.5194/angeo-22-3751-2004}

\bibitem[{Chandran {et~al.}(2011)Chandran, Dennis, Quataert, \&
  Bale}]{chandran_incorporating_2011}
Chandran, B. D.~G., Dennis, T.~J., Quataert, E., \& Bale, S.~D. 2011, The
  Astrophysical Journal, 743, 197, \dodoi{10.1088/0004-637X/743/2/197}

\bibitem[{Chandrasekhar {et~al.}(1958)Chandrasekhar, Kaufman, \&
  Watson}]{chandrasekhar_stability_1958}
Chandrasekhar, S., Kaufman, A., \& Watson, K. 1958, Proceedings of the Royal
  Society of London. Series A. Mathematical and Physical Sciences, 245, 435,
  \dodoi{10.1098/rspa.1958.0094}

\bibitem[{Chen(2016)}]{chen_recent_2016}
Chen, C. H.~K. 2016, Journal of Plasma Physics, 82,
  \dodoi{10.1017/S0022377816001124}

\bibitem[{Chen {et~al.}(2016)Chen, Matteini, Schekochihin, Stevens, Salem,
  Maruca, Kunz, \& Bale}]{chen_multi-species_2016}
Chen, C. H.~K., Matteini, L., Schekochihin, A.~A., {et~al.} 2016, The
  Astrophysical Journal, 825, L26, \dodoi{10.3847/2041-8205/825/2/L26}

\bibitem[{Chew {et~al.}(1956)Chew, Low, \& Goldberger}]{chew_boltzmann_1956}
Chew, G., Low, F., \& Goldberger, M. 1956, Proceedings of the Royal Society of
  London. Series A. Mathematical and Physical Sciences, 236, 112,
  \dodoi{10.1098/rspa.1956.0116}

\bibitem[{Cranmer {et~al.}(2015)Cranmer, Asgari-Targhi, Miralles, Raymond,
  Strachan, Tian, \& Woolsey}]{cranmer_role_2015}
Cranmer, S.~R., Asgari-Targhi, M., Miralles, M.~P., {et~al.} 2015,
  Philosophical Transactions of the Royal Society A: Mathematical, Physical and
  Engineering Sciences, 373, 20140148, \dodoi{10.1098/rsta.2014.0148}

\bibitem[{{D'Amicis} {et~al.}(2019){D'Amicis}, {De Marco}, {Bruno}, \&
  {Perrone}}]{damicis19}
{D'Amicis}, R., {De Marco}, R., {Bruno}, R., \& {Perrone}, D. 2019, \aap, 632,
  A92, \dodoi{10.1051/0004-6361/201936728}

\bibitem[{{D'Amicis} {et~al.}(2021){D'Amicis}, {Bruno}, {Panasenco}, {Telloni},
  {Perrone}, {Marcucci}, {Woodham}, {Velli}, {De Marco}, {Jagarlamudi}, {Coco},
  {Owen}, {Louarn}, {Livi}, {Horbury}, {Andr{\'e}}, {Angelini}, {Evans},
  {Fedorov}, {Genot}, {Lavraud}, {Matteini}, {M{\"u}ller}, {O'Brien}, {Pezzi},
  {Rouillard}, {Sorriso-Valvo}, {Tenerani}, {Verscharen}, \&
  {Zouganelis}}]{damicis_2021}
{D'Amicis}, R., {Bruno}, R., {Panasenco}, O., {et~al.} 2021, \aap, 656, A21,
  \dodoi{10.1051/0004-6361/202140938}

\bibitem[{Dong {et~al.}(2014)Dong, Verdini, \& Grappin}]{dong_evolution_2014}
Dong, Y., Verdini, A., \& Grappin, R. 2014, The Astrophysical Journal, 793,
  118, \dodoi{10.1088/0004-637X/793/2/118}

\bibitem[{Fox {et~al.}(2016)Fox, Velli, Bale, Decker, Driesman, Howard, Kasper,
  Kinnison, Kusterer, Lario, Lockwood, McComas, Raouafi, \&
  Szabo}]{fox_solar_2016}
Fox, N.~J., Velli, M.~C., Bale, S.~D., {et~al.} 2016, Space Science Reviews,
  204, 7, \dodoi{10.1007/s11214-015-0211-6}

\bibitem[{Franci {et~al.}(2015)Franci, Verdini, Matteini, Landi, \&
  Hellinger}]{franci_solar_2015}
Franci, L., Verdini, A., Matteini, L., Landi, S., \& Hellinger, P. 2015, The
  Astrophysical Journal, 804, L39, \dodoi{10.1088/2041-8205/804/2/l39}

\bibitem[{Gary(1993)}]{gary_theory_1993}
Gary, S.~P. 1993, Theory of space plasma microinstabilities, Cambridge
  atmospheric and space science series (Cambridge [England] ; New York:
  Cambridge University Press)

\bibitem[{Gary(2015)}]{gary_short-wavelength_2015}
---. 2015, Philosophical Transactions of the Royal Society A: Mathematical,
  Physical and Engineering Sciences, 373, 20140149,
  \dodoi{10.1098/rsta.2014.0149}

\bibitem[{Gary {et~al.}(2001)Gary, Skoug, Steinberg, \&
  Smith}]{gary_proton_2001}
Gary, S.~P., Skoug, R.~M., Steinberg, J.~T., \& Smith, C.~W. 2001, Geophysical
  Research Letters, 28, 2759, \dodoi{10.1029/2001GL013165}

\bibitem[{Hasegawa(1969)}]{hasegawa_drift_1969}
Hasegawa, A. 1969, Physics of Fluids, 12, 2642, \dodoi{10.1063/1.1692407}

\bibitem[{Hellinger {et~al.}(2017)Hellinger, Landi, Matteini, Verdini, \&
  Franci}]{hellinger_mirror_2017}
Hellinger, P., Landi, S., Matteini, L., Verdini, A., \& Franci, L. 2017, The
  Astrophysical Journal, 838, 158, \dodoi{10.3847/1538-4357/aa67e0}

\bibitem[{Hellinger {et~al.}(2015)Hellinger, Matteini, Landi, Verdini, Franci,
  \& Trávní\v{c}ek}]{hellinger_plasma_2015}
Hellinger, P., Matteini, L., Landi, S., {et~al.} 2015, The Astrophysical
  Journal, 811, L32, \dodoi{10.1088/2041-8205/811/2/L32}

\bibitem[{Hellinger {et~al.}(2006)Hellinger, Tr\'avn\v{c}ek, Kasper, \&
  Lazarus}]{hellinger_solar_2006}
Hellinger, P., Tr\'avn\v{c}ek, P., Kasper, J.~C., \& Lazarus, A.~J. 2006,
  Geophysical Research Letters, 33, \dodoi{10.1029/2006GL025925}

\bibitem[{Hellinger {et~al.}(2013)Hellinger, Tr\'vní\v{c}ek, Štverák,
  Matteini, \& Velli}]{hellinger_proton_2013}
Hellinger, P., Tr\'vní\v{c}ek, P.~M., Štverák, v., Matteini, L., \& Velli,
  M. 2013, Journal of Geophysical Research: Space Physics, 118, 1351,
  \dodoi{10.1002/jgra.50107}

\bibitem[{Hellinger \& Trávníček(2008)}]{hellinger_oblique_2008}
Hellinger, P., \& Trávníček, P.~M. 2008, Journal of Geophysical Research:
  Space Physics, 113, \dodoi{10.1029/2008JA013416}

\bibitem[{{Horbury} {et~al.}(2020){Horbury}, {O'Brien}, {Carrasco Blazquez},
  {Bendyk}, {Brown}, {Hudson}, {Evans}, {Oddy}, {Carr}, {Beek}, {Cupido},
  {Bhattacharya}, {Dominguez}, {Matthews}, {Myklebust}, {Whiteside}, {Bale},
  {Baumjohann}, {Burgess}, {Carbone}, {Cargill}, {Eastwood}, {Erd{\"o}s},
  {Fletcher}, {Forsyth}, {Giacalone}, {Glassmeier}, {Goldstein}, {Hoeksema},
  {Lockwood}, {Magnes}, {Maksimovic}, {Marsch}, {Matthaeus}, {Murphy},
  {Nakariakov}, {Owen}, {Owens}, {Rodriguez-Pacheco}, {Richter}, {Riley},
  {Russell}, {Schwartz}, {Vainio}, {Velli}, {Vennerstrom}, {Walsh},
  {Wimmer-Schweingruber}, {Zank}, {M{\"u}ller}, {Zouganelis}, \&
  {Walsh}}]{horburymag20}
{Horbury}, T.~S., {O'Brien}, H., {Carrasco Blazquez}, I., {et~al.} 2020, \aap,
  642, A9, \dodoi{10.1051/0004-6361/201937257}

\bibitem[{Howes(2015)}]{howes_dynamical_2015}
Howes, G.~G. 2015, Philosophical Transactions of the Royal Society A:
  Mathematical, Physical and Engineering Sciences, 373, 20140145,
  \dodoi{10.1098/rsta.2014.0145}

\bibitem[{Howes {et~al.}(2006)Howes, Cowley, Dorland, Hammett, Quataert, \&
  Schekochihin}]{howes_astrophysical_2006}
Howes, G.~G., Cowley, S.~C., Dorland, W., {et~al.} 2006, The Astrophysical
  Journal, 651, 590, \dodoi{10.1086/506172}

\bibitem[{Howes {et~al.}(2008)Howes, Cowley, Dorland, Hammett, Quataert, \&
  Schekochihin}]{howes_model_2008}
---. 2008, Journal of Geophysical Research: Space Physics, 113

\bibitem[{Howes {et~al.}(2011)Howes, TenBarge, Dorland, Quataert, Schekochihin,
  Numata, \& Tatsuno}]{howes_gyrokinetic_2011}
Howes, G.~G., TenBarge, J.~M., Dorland, W., {et~al.} 2011, Physical Review
  Letters, 107, 035004, \dodoi{10.1103/PhysRevLett.107.035004}

\bibitem[{Isenberg {et~al.}(2013)Isenberg, Maruca, \&
  Kasper}]{isenberg_self-consistent_2013}
Isenberg, P.~A., Maruca, B.~A., \& Kasper, J.~C. 2013, The Astrophysical
  Journal, 773, 164, \dodoi{10.1088/0004-637X/773/2/164}

\bibitem[{Kasper {et~al.}(2002)Kasper, Lazarus, \& Gary}]{kasper_windswe_2002}
Kasper, J.~C., Lazarus, A.~J., \& Gary, S.~P. 2002, Geophysical Research
  Letters, 29, 20, \dodoi{10.1029/2002GL015128}

\bibitem[{{Kivelson} \& {Southwood}(1996)}]{kivelson1996}
{Kivelson}, M.~G., \& {Southwood}, D.~J. 1996, \jgr, 101, 17365,
  \dodoi{10.1029/96JA01407}

\bibitem[{Kiyani {et~al.}(2015)Kiyani, Osman, \&
  Chapman}]{kiyani_dissipation_2015}
Kiyani, K.~H., Osman, K.~T., \& Chapman, S.~C. 2015, Philosophical Transactions
  of the Royal Society A: Mathematical, Physical and Engineering Sciences, 373,
  20140155, \dodoi{10.1098/rsta.2014.0155}

\bibitem[{Klein {et~al.}(2018)Klein, Alterman, Stevens, Vech, \&
  Kasper}]{klein_majority_2018}
Klein, K., Alterman, B., Stevens, M., Vech, D., \& Kasper, J. 2018, Physical
  Review Letters, 120, 205102, \dodoi{10.1103/PhysRevLett.120.205102}

\bibitem[{Klein {et~al.}(2017)Klein, Kasper, Korreck, \&
  Stevens}]{klein_applying_2017}
Klein, K.~G., Kasper, J.~C., Korreck, K.~E., \& Stevens, M.~L. 2017, Journal of
  Geophysical Research: Space Physics, 122, 9815, \dodoi{10.1002/2017JA024486}

\bibitem[{{Kunz} {et~al.}(2015){Kunz}, {Schekochihin}, {Chen}, {Abel}, \&
  {Cowley}}]{kunz2015}
{Kunz}, M.~W., {Schekochihin}, A.~A., {Chen}, C.~H.~K., {Abel}, I.~G., \&
  {Cowley}, S.~C. 2015, Journal of Plasma Physics, 81, 325810501,
  \dodoi{10.1017/S0022377815000811}

\bibitem[{Kunz {et~al.}(2014)Kunz, Schekochihin, \& Stone}]{kunz_firehose_2014}
Kunz, M.~W., Schekochihin, A.~A., \& Stone, J.~M. 2014, Physical Review
  Letters, 112, 205003, \dodoi{10.1103/PhysRevLett.112.205003}

\bibitem[{{Kunz} {et~al.}(2016){Kunz}, {Stone}, \& {Quataert}}]{Kunz2016}
{Kunz}, M.~W., {Stone}, J.~M., \& {Quataert}, E. 2016, \prl, 117, 235101,
  \dodoi{10.1103/PhysRevLett.117.235101}

\bibitem[{{Louarn} {et~al.}(2021){Louarn}, {Fedorov}, {Prech}, {Owen}, {Bruno},
  {Livi}, {Lavraud}, {Rouillard}, {G{\'e}not}, {Andr{\'e}}, {Fruit},
  {R{\'e}ville}, {Kieokaew}, {Plotnikov}, {Penou}, {Barthe}, {Khataria},
  {Berthomier}, {D'Amicis}, {Sorriso-Valvo}, {Allegrini}, {Raines},
  {Verscharen}, {Fortunato}, {Mele}, {Horbury}, {O'brien}, {Evans}, {Angelini},
  {Maksimovic}, {Kasper}, \& {Bale}}]{louarn21}
{Louarn}, P., {Fedorov}, A., {Prech}, L., {et~al.} 2021, \aap, 656, A36,
  \dodoi{10.1051/0004-6361/202141095}

\bibitem[{López {et~al.}(2022)López, Micera, Lazar, Poedts, Lapenta, Zhukov,
  Boella, \& Shaaban}]{lopez_mixing_2022}
López, R.~A., Micera, A., Lazar, M., {et~al.} 2022, The Astrophysical Journal,
  930, 158, \dodoi{10.3847/1538-4357/ac66e4}

\bibitem[{Marsch {et~al.}(2004)Marsch, Ao, \& Tu}]{marsch_temperature_2004}
Marsch, E., Ao, X.-Z., \& Tu, C.-Y. 2004, Journal of Geophysical Research, 109,
  A04102, \dodoi{10.1029/2003JA010330}

\bibitem[{{Maruca} \& {Kasper}(2013)}]{Maruca2013}
{Maruca}, B.~A., \& {Kasper}, J.~C. 2013, Advances in Space Research, 52, 723,
  \dodoi{10.1016/j.asr.2013.04.006}

\bibitem[{{Maruca} {et~al.}(2012){Maruca}, {Kasper}, \& {Gary}}]{maruca+2012}
{Maruca}, B.~A., {Kasper}, J.~C., \& {Gary}, S.~P. 2012, \apj, 748, 137,
  \dodoi{10.1088/0004-637X/748/2/137}

\bibitem[{Matteini {et~al.}(2012)Matteini, Hellinger, Landi, Trávníček, \&
  Velli}]{matteini_ion_2012}
Matteini, L., Hellinger, P., Landi, S., Trávníček, P.~M., \& Velli, M. 2012,
  Space Science Reviews, 172, 373, \dodoi{10.1007/s11214-011-9774-z}

\bibitem[{Matteini {et~al.}(2007)Matteini, Landi, Hellinger, Pantellini,
  Maksimovic, Velli, Goldstein, \& Marsch}]{matteini_evolution_2007}
Matteini, L., Landi, S., Hellinger, P., {et~al.} 2007, Geophysical Research
  Letters, 34, L20105, \dodoi{10.1029/2007GL030920}

\bibitem[{Matteini {et~al.}(2006)Matteini, Landi, Hellinger, \&
  Velli}]{matteini_parallel_2006}
Matteini, L., Landi, S., Hellinger, P., \& Velli, M. 2006, Journal of
  Geophysical Research: Space Physics, 111,
  \dodoi{https://doi.org/10.1029/2006JA011667}

\bibitem[{Matthaeus {et~al.}(2005)Matthaeus, Dasso, Weygand, Milano, Smith, \&
  Kivelson}]{matthaeus_spatial_2005}
Matthaeus, W.~H., Dasso, S., Weygand, J.~M., {et~al.} 2005, Physical Review
  Letters, 95, 231101, \dodoi{10.1103/PhysRevLett.95.231101}

\bibitem[{Matthaeus \& Goldstein(1982)}]{matthaeus_measurement_1982}
Matthaeus, W.~H., \& Goldstein, M.~L. 1982, Journal of Geophysical Research,
  87, 6011, \dodoi{10.1029/JA087iA08p06011}

\bibitem[{Matthaeus {et~al.}(2014)Matthaeus, Oughton, Osman, Servidio, Wan,
  Gary, Shay, Valentini, Roytershteyn, Karimabadi, \&
  Chapman}]{matthaeus_nonlinear_2014}
Matthaeus, W.~H., Oughton, S., Osman, K.~T., {et~al.} 2014, The Astrophysical
  Journal, 790, 155, \dodoi{10.1088/0004-637X/790/2/155}

\bibitem[{{M{\"u}ller} {et~al.}(2020){M{\"u}ller}, {St. Cyr}, {Zouganelis},
  {Gilbert}, {Marsden}, {Nieves-Chinchilla}, {Antonucci}, {Auch{\`e}re},
  {Berghmans}, {Horbury}, {Howard}, {Krucker}, {Maksimovic}, {Owen}, {Rochus},
  {Rodriguez-Pacheco}, {Romoli}, {Solanki}, {Bruno}, {Carlsson}, {Fludra},
  {Harra}, {Hassler}, {Livi}, {Louarn}, {Peter}, {Sch{\"u}hle}, {Teriaca}, {del
  Toro Iniesta}, {Wimmer-Schweingruber}, {Marsch}, {Velli}, {De Groof},
  {Walsh}, \& {Williams}}]{mueller20}
{M{\"u}ller}, D., {St. Cyr}, O.~C., {Zouganelis}, I., {et~al.} 2020, \aap, 642,
  A1, \dodoi{10.1051/0004-6361/202038467}

\bibitem[{Nicolaou {et~al.}(2019)Nicolaou, Verscharen, Wicks, \&
  Owen}]{nicolaou_impact_2019}
Nicolaou, G., Verscharen, D., Wicks, R.~T., \& Owen, C.~J. 2019, The
  Astrophysical Journal, 886, 101, \dodoi{10.3847/1538-4357/ab48e3}

\bibitem[{{Nicolaou} {et~al.}(2021){Nicolaou}, {Wicks}, {Owen}, {Kataria},
  {Chandrasekhar}, {Lewis}, {Verscharen}, {Fortunato}, {Mele}, {DeMarco}, \&
  {Bruno}}]{nicolaou21orbiter}
{Nicolaou}, G., {Wicks}, R.~T., {Owen}, C.~J., {et~al.} 2021, \aap, 656, A10,
  \dodoi{10.1051/0004-6361/202140875}

\bibitem[{Ogilvie {et~al.}(1995)Ogilvie, Chornay, Fritzenreiter, Hunsaker,
  Keller, Lobell, Miller, Scudder, Sittler, Torbert, Bodet, Needell, Lazarus,
  Steinberg, Tappan, Mavretic, \& Gergin}]{ogilvie_swe_1995}
Ogilvie, K.~W., Chornay, D.~J., Fritzenreiter, R.~J., {et~al.} 1995, Space
  Science Reviews, 71, 55, \dodoi{10.1007/BF00751326}

\bibitem[{{Owen} {et~al.}(2020){Owen}, {Bruno}, {Livi}, {Louarn}, {Al Janabi},
  {Allegrini}, {Amoros}, {Baruah}, {Barthe}, {Berthomier}, {Bordon},
  {Brockley-Blatt}, {Brysbaert}, {Capuano}, {Collier}, {DeMarco}, {Fedorov},
  {Ford}, {Fortunato}, {Fratter}, {Galvin}, {Hancock}, {Heirtzler}, {Kataria},
  {Kistler}, {Lepri}, {Lewis}, {Loeffler}, {Marty}, {Mathon}, {Mayall}, {Mele},
  {Ogasawara}, {Orlandi}, {Pacros}, {Penou}, {Persyn}, {Petiot}, {Phillips},
  {P{\v{r}}ech}, {Raines}, {Reden}, {Rouillard}, {Rousseau}, {Rubiella},
  {Seran}, {Spencer}, {Thomas}, {Trevino}, {Verscharen}, {Wurz}, {Alapide},
  {Amoruso}, {Andr{\'e}}, {Anekallu}, {Arciuli}, {Arnett}, {Ascolese},
  {Bancroft}, {Bland}, {Brysch}, {Calvanese}, {Castronuovo},
  {{\v{C}}erm{\'a}k}, {Chornay}, {Clemens}, {Coker}, {Collinson}, {D'Amicis},
  {Dandouras}, {Darnley}, {Davies}, {Davison}, {De Los Santos}, {Devoto},
  {Dirks}, {Edlund}, {Fazakerley}, {Ferris}, {Frost}, {Fruit}, {Garat},
  {G{\'e}not}, {Gibson}, {Gilbert}, {de Giosa}, {Gradone}, {Hailey}, {Horbury},
  {Hunt}, {Jacquey}, {Johnson}, {Lavraud}, {Lawrenson}, {Leblanc}, {Lockhart},
  {Maksimovic}, {Malpus}, {Marcucci}, {Mazelle}, {Monti}, {Myers}, {Nguyen},
  {Rodriguez-Pacheco}, {Phillips}, {Popecki}, {Rees}, {Rogacki}, {Ruane},
  {Rust}, {Salatti}, {Sauvaud}, {Stakhiv}, {Stange}, {Stubbs}, {Taylor},
  {Techer}, {Terrier}, {Thibodeaux}, {Urdiales}, {Varsani}, {Walsh}, {Watson},
  {Wheeler}, {Willis}, {Wimmer-Schweingruber}, {Winter}, {Yardley}, \&
  {Zouganelis}}]{owen20}
{Owen}, C.~J., {Bruno}, R., {Livi}, S., {et~al.} 2020, \aap, 642, A16,
  \dodoi{10.1051/0004-6361/201937259}

\bibitem[{Owen {et~al.}(2021)Owen, Foster, Bruno, Livi, Louarn, Berthomier,
  Fedorov, Anekallu, Kataria, Kelly, Lewis, Watson, Berčič, Stansby, Suen,
  Verscharen, Fortunato, Nicolaou, Wicks, Rae, Lavraud, Horbury, O’Brien,
  Evans, \& Angelini}]{owen_solar_2021}
Owen, C.~J., Foster, A.~C., Bruno, R., {et~al.} 2021, Astronomy \&
  Astrophysics, 656, L8, \dodoi{10.1051/0004-6361/202140944}

\bibitem[{Parker(1965)}]{parker_dynamical_1965}
Parker, E. 1965, Space Science Reviews, 4, 666

\bibitem[{Parker(1958)}]{parker_dynamical_1958}
Parker, E.~N. 1958, Physical Review, 109, 1874,
  \dodoi{10.1103/PhysRev.109.1874}

\bibitem[{Pokhotelov(2004)}]{pokhotelov_mirror_2004}
Pokhotelov, O.~A. 2004, Journal of Geophysical Research, 109, A09213,
  \dodoi{10.1029/2004JA010568}

\bibitem[{Qudsi {et~al.}(2020)Qudsi, Maruca, Matthaeus, Parashar,
  Bandyopadhyay, Chhiber, Chasapis, Goldstein, Bale, Bonnell, Wit, Goetz,
  Harvey, MacDowall, Malaspina, Pulupa, Kasper, Korreck, Case, Stevens,
  Whittlesey, Larson, Livi, Velli, \& Raouafi}]{qudsi_observations_2020}
Qudsi, R.~A., Maruca, B.~A., Matthaeus, W.~H., {et~al.} 2020, The Astrophysical
  Journal Supplement Series, 246, 46, \dodoi{10.3847/1538-4365/ab5c19}

\bibitem[{Russell {et~al.}(1999)Russell, Huddleston, Strangeway, Blanco-Cano,
  Kivelson, Khurana, Frank, Paterson, Gurnett, \&
  Kurth}]{russell_mirror-mode_1999}
Russell, C.~T., Huddleston, D.~E., Strangeway, R.~J., {et~al.} 1999, Journal of
  Geophysical Research: Space Physics, 104, 17471, \dodoi{10.1029/1999JA900202}

\bibitem[{Schekochihin {et~al.}(2009)Schekochihin, Cowley, Dorland, Hammett,
  Howes, Quataert, \& Tatsuno}]{schekochihin_astrophysical_2009}
Schekochihin, A.~A., Cowley, S.~C., Dorland, W., {et~al.} 2009, The
  Astrophysical Journal Supplement Series, 182, 310,
  \dodoi{10.1088/0067-0049/182/1/310}

\bibitem[{Servidio {et~al.}(2014)Servidio, Osman, Valentini, Perrone, Califano,
  Chapman, Matthaeus, \& Veltri}]{servidio_proton_2014}
Servidio, S., Osman, K.~T., Valentini, F., {et~al.} 2014, The Astrophysical
  Journal, 781, L27, \dodoi{10.1088/2041-8205/781/2/L27}

\bibitem[{Taylor(1938)}]{taylor_spectrum_1938}
Taylor, G.~I. 1938, Proceedings of the Royal Society A: Mathematical, Physical
  and Engineering Sciences, 164, 476, \dodoi{10.1098/rspa.1938.0032}

\bibitem[{Treumann {et~al.}(2019)Treumann, Baumjohann, \&
  Narita}]{treumann_applicability_2019}
Treumann, R.~A., Baumjohann, W., \& Narita, Y. 2019, Earth, Planets and Space,
  71, 41, \dodoi{10.1186/s40623-019-1021-y}

\bibitem[{Tu \& Marsch(1995)}]{tu_mhd_1995}
Tu, C.-Y., \& Marsch, E. 1995, Space Science Reviews, 73, 1

\bibitem[{Verscharen {et~al.}(2016)Verscharen, Chandran, Klein, \&
  Quataert}]{verscharen16}
Verscharen, D., Chandran, B. D.~G., Klein, K.~G., \& Quataert, E. 2016, The
  Astrophysical Journal, 831, 128, \dodoi{10.3847/0004-637X/831/2/128}

\bibitem[{{Verscharen} {et~al.}(2017){Verscharen}, {Chen}, \&
  {Wicks}}]{verscharen17}
{Verscharen}, D., {Chen}, C. H.~K., \& {Wicks}, R.~T. 2017, \apj, 840, 106,
  \dodoi{10.3847/1538-4357/aa6a56}

\bibitem[{Verscharen {et~al.}(2019)Verscharen, Klein, \&
  Maruca}]{verscharen_multi-scale_2019}
Verscharen, D., Klein, K.~G., \& Maruca, B.~A. 2019, Living Reviews in Solar
  Physics, 16, \dodoi{10.1007/s41116-019-0021-0}

\bibitem[{Verscharen \& Marsch(2011)}]{verscharen_apparent_2011}
Verscharen, D., \& Marsch, E. 2011, Annales Geophysicae, 29, 909,
  \dodoi{10.5194/angeo-29-909-2011}

\bibitem[{{Wicks} {et~al.}(2016){Wicks}, {Alexander}, {Stevens}, {Wilson},
  {Moya}, {Vi{\~n}as}, {Jian}, {Roberts}, {O'Modhrain}, {Gilbert}, \&
  {Zurbuchen}}]{wicks2016}
{Wicks}, R.~T., {Alexander}, R.~L., {Stevens}, M., {et~al.} 2016, \apj, 819, 6,
  \dodoi{10.3847/0004-637X/819/1/6}

\bibitem[{{Woodham} {et~al.}(2021){Woodham}, {Horbury}, {Matteini}, {Woolley},
  {Laker}, {Bale}, {Nicolaou}, {Stawarz}, {Stansby}, {Hietala}, {Larson},
  {Livi}, {Verniero}, {McManus}, {Kasper}, {Korreck}, {Raouafi}, {Moncuquet},
  \& {Pulupa}}]{woodham21}
{Woodham}, L.~D., {Horbury}, T.~S., {Matteini}, L., {et~al.} 2021, \aap, 650,
  L1, \dodoi{10.1051/0004-6361/202039415}

\bibitem[{{Yoon}(2016)}]{yoon2016}
{Yoon}, P.~H. 2016, The Astrophysical Journal, 833, 106,
  \dodoi{10.3847/1538-4357/833/1/106}

\bibitem[{Yoon {et~al.}(2021)Yoon, Sarfraz, Ali, Salem, \&
  Seough}]{yoon_proton_2021}
Yoon, P.~H., Sarfraz, M., Ali, Z., Salem, C.~S., \& Seough, J. 2021, Monthly
  Notices of the Royal Astronomical Society, 509, 4736,
  \dodoi{10.1093/mnras/stab3286}

\bibitem[{{Zouganelis} {et~al.}(2020){Zouganelis}, {De Groof}, {Walsh},
  {Williams}, {M{\"u}ller}, {St Cyr}, {Auch{\`e}re}, {Berghmans}, {Fludra},
  {Horbury}, {Howard}, {Krucker}, {Maksimovic}, {Owen},
  {Rodr{\'\i}guez-Pacheco}, {Romoli}, {Solanki}, {Watson}, {Sanchez}, {Lefort},
  {Osuna}, {Gilbert}, {Nieves-Chinchilla}, {Abbo}, {Alexandrova},
  {Anastasiadis}, {Andretta}, {Antonucci}, {Appourchaux}, {Aran}, {Arge},
  {Aulanier}, {Baker}, {Bale}, {Battaglia}, {Bellot Rubio}, {Bemporad},
  {Berthomier}, {Bocchialini}, {Bonnin}, {Brun}, {Bruno}, {Buchlin},
  {B{\"u}chner}, {Bucik}, {Carcaboso}, {Carr}, {Carrasco-Bl{\'a}zquez},
  {Cecconi}, {Cernuda Cangas}, {Chen}, {Chitta}, {Chust}, {Dalmasse},
  {D'Amicis}, {Da Deppo}, {De Marco}, {Dolei}, {Dolla}, {Dudok de Wit}, {van
  Driel-Gesztelyi}, {Eastwood}, {Espinosa Lara}, {Etesi}, {Fedorov},
  {F{\'e}lix-Redondo}, {Fineschi}, {Fleck}, {Fontaine}, {Fox}, {Gandorfer},
  {G{\'e}not}, {Georgoulis}, {Gissot}, {Giunta}, {Gizon}, {G{\'o}mez-Herrero},
  {Gontikakis}, {Graham}, {Green}, {Grundy}, {Haberreiter}, {Harra}, {Hassler},
  {Hirzberger}, {Ho}, {Hurford}, {Innes}, {Issautier}, {James}, {Janitzek},
  {Janvier}, {Jeffrey}, {Jenkins}, {Khotyaintsev}, {Klein}, {Kontar},
  {Kontogiannis}, {Krafft}, {Krasnoselskikh}, {Kretzschmar}, {Labrosse},
  {Lagg}, {Landini}, {Lavraud}, {Leon}, {Lepri}, {Lewis}, {Liewer}, {Linker},
  {Livi}, {Long}, {Louarn}, {Malandraki}, {Maloney}, {Martinez-Pillet},
  {Martinovic}, {Masson}, {Matthews}, {Matteini}, {Meyer-Vernet}, {Moraitis},
  {Morton}, {Musset}, {Nicolaou}, {Nindos}, {O'Brien}, {Orozco Suarez},
  {Owens}, {Pancrazzi}, {Papaioannou}, {Parenti}, {Pariat}, {Patsourakos},
  {Perrone}, {Peter}, {Pinto}, {Plainaki}, {Plettemeier}, {Plunkett}, {Raines},
  {Raouafi}, {Reid}, {Retino}, {Rezeau}, {Rochus}, {Rodriguez},
  {Rodriguez-Garcia}, {Roth}, {Rouillard}, {Sahraoui}, {Sasso}, {Schou},
  {Sch{\"u}hle}, {Sorriso-Valvo}, {Soucek}, {Spadaro}, {Stangalini}, {Stansby},
  {Steller}, {Strugarek}, {{\v{S}}tver{\'a}k}, {Susino}, {Telloni}, {Terasa},
  {Teriaca}, {Toledo-Redondo}, {del Toro Iniesta}, {Tsiropoula}, {Tsounis},
  {Tziotziou}, {Valentini}, {Vaivads}, {Vecchio}, {Velli}, {Verbeeck},
  {Verdini}, {Verscharen}, {Vilmer}, {Vourlidas}, {Wicks},
  {Wimmer-Schweingruber}, {Wiegelmann}, {Young}, \& {Zhukov}}]{zouganelis2020}
{Zouganelis}, I., {De Groof}, A., {Walsh}, A.~P., {et~al.} 2020, \aap, 642, A3,
  \dodoi{10.1051/0004-6361/202038445}

\end{thebibliography}
\bibliographystyle{aasjournal}



\end{document}